\documentstyle[12pt]{article}
\input psfig.sty
\def\nn{\noindent}
\def\Re{{\cal R \mskip-4mu \lower.1ex \hbox{\it e}\,}}
\def\Im{{\cal I \mskip-5mu \lower.1ex \hbox{\it m}\,}}
\def\ie{{\it i.e.}}
\def\eg{{\it e.g.}}
\def\etc{{\it etc}}
\def\etal{{\it et al.}}

\def\sub#1{_{\lower.25ex\hbox{$\scriptstyle#1$}}}
\def\tev{\,{\ifmmode\mathrm {TeV}\else TeV\fi}}
\def\gev{\,{\ifmmode\mathrm {GeV}\else GeV\fi}}
\def\mev{\,{\ifmmode\mathrm {MeV}\else MeV\fi}}
\def\to{\rightarrow}

\def\subw{_{\rm w}}
\def\mh{\ifmmode m\sbl H \else $m\sbl H$\fi}
\def\mch{\ifmmode m_{H^\pm} \else $m_{H^\pm}$\fi}
\def\mt{\ifmmode m_t\else $m_t$\fi}
\def\mc{\ifmmode m_c\else $m_c$\fi}
\def\mz{\ifmmode M_Z\else $M_Z$\fi}
\def\mw{\ifmmode M_W\else $M_W$\fi}
\def\mws{\ifmmode M_W^2 \else $M_W^2$\fi}
\def\mhs{\ifmmode m_H^2 \else $m_H^2$\fi}   
\def\mzs{\ifmmode M_Z^2 \else $M_Z^2$\fi}
\def\mts{\ifmmode m_t^2 \else $m_t^2$\fi}
\def\mcs{\ifmmode m_c^2 \else $m_c^2$\fi}
\def\mchs{\ifmmode m_{H^\pm}^2 \else $m_{H^\pm}^2$\fi}
\def\ztwo{\ifmmode Z_2\else $Z_2$\fi}
\def\zone{\ifmmode Z_1\else $Z_1$\fi}
\def\mtwo{\ifmmode M_2\else $M_2$\fi}
\def\mone{\ifmmode M_1\else $M_1$\fi}
\def\tb{\ifmmode \tan\beta \else $\tan\beta$\fi}
\def\xw{\ifmmode x\subw\else $x\subw$\fi}
\def\ch{\ifmmode H^\pm \else $H^\pm$\fi}
\def\lum{\ifmmode {\cal L}\else ${\cal L}$\fi}
\def\inpb{\,{\ifmmode {\mathrm {pb}}^{-1}\else ${\mathrm {pb}}^{-1}$\fi}}
\def\infb{\,{\ifmmode {\mathrm {fb}}^{-1}\else ${\mathrm {fb}}^{-1}$\fi}}
\def\epem{\ifmmode e^+e^-\else $e^+e^-$\fi}
\def\ppb{\ifmmode \bar pp\else $\bar pp$\fi}
\def\bsg{\ifmmode B\to X_s\gamma\else $B\to X_s\gamma$\fi}
\def\bsll{\ifmmode B\to X_s\ell^+\ell^-\else $B\to X_s\ell^+\ell^-$\fi}
\def\bstt{\ifmmode B\to X_s\tau^+\tau^-\else $B\to X_s\tau^+\tau^-$\fi}
\def\lamt{\ifmmode \tilde\lambda\else $\tilde\lambda$\fi}
\def\shat{\ifmmode \hat s\else $\hat s$\fi}
\def\that{\ifmmode \hat t\else $\hat t$\fi}
\def\uhat{\ifmmode \hat u\else $\hat u$\fi}

\newskip\zatskip \zatskip=0pt plus0pt minus0pt
\def\matth{\mathsurround=0pt}

\def\atversim#1#2{\lower0.7ex\vbox{\baselineskip\zatskip\lineskip\zatskip
  \lineskiplimit 0pt\ialign{$\matth#1\hfil##\hfil$\crcr#2\crcr\sim\crcr}}}

\renewcommand{\thefootnote}{\fnsymbol{footnote}}

\begin{document} \begin{titlepage} 
\rightline{\vbox{\halign{&#\hfil\cr
&SLAC-PUB-7738\cr
&March 1998\cr}}}
\begin{center}

{\Large\bf Right-Handed Currents in B Decay Revisited}
\footnote{Work supported by the Department of 
Energy, Contract DE-AC03-76SF00515}
\medskip

\normalsize 
{\large Thomas G. Rizzo } \\
\vskip .3cm
Stanford Linear Accelerator Center \\
Stanford CA 94309, USA\\
\vskip .3cm

\end{center}

\begin{abstract} 
 
We critically re-examine the case for and against a sizeable right-handed 
component in the $b\to c$ charged current coupling with a strength $\xi$ 
relative to the conventional left-handed current. Using data from CLEO on the 
decay $B\to D^*\ell \nu$, as well as our knowledge of $V_{cb}$ extracted from 
both inclusive and exclusive processes, we are able to determine the presently 
allowed parameter space for $\xi$ via HQET. We then identify several 
observables which could be measured at $B$ factories to either strengthen 
these constraints or otherwise observe right-handed currents. This parameter 
space region is found to be consistent with the low degree of $\Lambda_b$ 
polarization as determined by ALEPH as well as the measurements of the charged 
lepton and neutrino energy spectra from $b$ decay made by L3. We discuss how 
future measurements of semileptonic decay distributions may distinguish 
between exotic $\Lambda_b$ depolarization mechanisms and the existence of 
right-handed currents. Within the parameter space allowed by CLEO, using the 
Left-Right Symmetric Model as a guide, we perform a detailed search for 
specific sub-regions which can lead to a reduction in both the $B$ 
semileptonic branching fraction as well as the the average yield of charmed 
quarks in $B$ decay. The results provide a concrete realization of an earlier 
suggestion by Voloshin but may lead to potential difficulties with certain 
penguin mediated decay processes. 

\end{abstract} 




\renewcommand{\thefootnote}{\arabic{footnote}} \end{titlepage}


\section{Introduction}

The new generation of $B$ factories soon to come on line will open up an 
additional window on potential physics beyond the Standard Model(SM). Unlike 
the situation at new higher energy colliders, the physics beyond the SM 
will appear indirectly, \eg, as deviations from SM expectations in 
decay rates, distributions and/or asymmetries obtained through precision 
measurements. Such measurements may be as important in probing the SM as are 
those currently performed by LEP/SLC and the Tevatron at higher energies and 
will rival others associated with the observation of CP violation in the $B$ 
system. 

Perhaps the most fundamental of all quantities associated with $b$ 
quark decay is the chirality of its charged current coupling. 
The possibility that the $b\to c$ charged current(CC) may have a sizeable 
right-handed(RH) component has been the subject of speculation for some time. 
Early on, Gronau and Wakaizumi, as well as a number of other 
authors{\cite {gro}}, speculated that the $b\to c$ coupling might be almost, 
if not {\it purely}, RH. Thanks to measurements performed by both the 
CLEO{\cite {cleo}} and L3{\cite {l3}} Collaborations, to which we return 
at some length below, we now know that this hypothesis cannot be true. 
The relative 
strength of the RH $b\to c$ coupling in comparison to the corresponding SM 
left-handed(LH) coupling must be somewhat less than unity; the leptonic 
current in the decay is highly constrained to be LH. It is important to 
observe that the 
results of these experiments cannot exclude a RH coupling of modest strength. 
Other data, such as the apparently small value of the $\Lambda_b$ polarization 
observed in $Z$ decay{\cite {lambda}} by ALEPH, qualitatively support the 
hypothesis of potentially sizeable RH couplings unless some exotic 
depolarization mechanisms are at work. Thus the current experimental situation 
remains unsatisfying and is far from resolving the issue of the presence 
of RH couplings in $b\to c$ transitions. On the theoretical side, in a 
completely different 
context, Voloshin{\cite {vol}} has recently suggested that a RH $b\to c$ 
coupling of modest strength may help to resolve the well-known{cite {blok}} 
$B$ semileptonic branching fraction($B_\ell$) and charm counting($n_c$) 
problems{\cite {prob}}. Thus we are left with the questions: given the present 
data, it is possible that such a RH coupling actually exists and can it 
assist with the $B_\ell$ and $n_c$ problems?  How can future $B$ factory 
data help clarify this situation? 

In this paper we will examine the simultaneous compatibility of the CLEO and, 
to a lesser extent, the L3 constraints on the RH $b\to c$ coupling and the 
desire to address the charm counting/branching fraction 
problem along the lines suggested by Voloshin. The important role played by the 
ALEPH $\Lambda_b$ polarization measurement is also examined. This discussion 
stresses both what we can learn from the current data about possible $b\to c$ 
RH couplings and what can be learned through future precision 
measurements at $B$ 
factories to resolve the present ambiguous situation. Most of this 
analysis, associated with the constraints from the CLEO, L3 and ALEPH data, 
can be performed in a completely {\it model-independent} fashion without any 
reference as to the possible origin of the $b\to c$ RH coupling. However, in 
order to subsequently approach the $n_c-B_\ell$ problem a more rigid 
theoretical framework, such as the Left-Right Symmetric 
Model(LRM){\cite {lrm}}, needs to be invoked. Other more general frameworks 
are possible but are beyond the scope of the present paper.

This paper is organized as follows. In Section 2, we re-examine and update 
the constraints imposed on the relative strength of the RH to LH $b\to c$ 
coupling due to current experimental data from CLEO, ALEPH and L3 using a 
model-independent approach and relying on Heavy Quark Effective Theory. As 
we will see there is a tantalizing, though still not compelling, hint of RH 
interactions in the CLEO data. 
The importance of both the small observed $\Lambda_b$ polarization as a 
possible, though still ambiguous, signature for RH currents is discussed. We 
also examine how this scenario may be distinguished from the SM with exotic 
depolarization mechanisms. The present ALEPH data is shown to be consistent 
with a moderately strong RH current coupling. In examining how we can extract 
further information from the present data and 
with an eye towards future measurements we discuss several new observables and 
their potential usefulness in probing for RH couplings. Many of these 
observables have either not been measured or have yet to be examined with 
any degree of precision. Several of these quantities can be probed at the 
$Z$ or during the first year of running at the new $B$ factories. In 
Section 3, we present an overview of the LRM and discuss the meaning of the 
current experimental constraints within this specific context keeping detailed 
discussion of the required LRM particle 
content to a minimum. As we do not want to restrict or constrain 
ourselves to a specific version of this model, we tacitly avoid at this 
point any discussion of loop processes which may involve the full particle 
spectrum of a realistic, probably supersymmetric LRM. In Section 4 we 
describe the nonleptonic $b\to c$ 
decays in the presence of RH currents and their associated decay widths 
including LO and estimates of the NLO QCD corrections based upon what is 
currently known in the case of the SM. In Section 5, we use the LRM as input 
to scan the full model parameter space allowed by the CLEO data to discover and 
identify sub-regions that will lead to a simultaneously decrease in the 
values of both $B_\ell$ and $n_c$ in comparison to the SM expectations. While 
such regions are shown to exist they occur in only a small, fine-tuned, 
fraction of the entire parameter space volume. 
Our summary and conclusions can be found in Section 6. In the Appendix we 
speculate on the possible forms of $V_R$ and point 
out how certain penguin processes can lead to difficulties with the solutions 
to the $n_c-B_\ell$ problem that we've obtained.

\section{Constraints on Right-Handed $b\to c$ Couplings}

\subsection{Model-Independent Notation}

Allowing for both LH and RH $b\to c$ couplings while, following Voloshin, 
maintaining the leptonic current as purely LH to satisfy the well-known 
$\mu$ decay constraints{\cite {muon}} without fine-tuning neutrino masses, the 
general four-fermion interaction describing $B$ semileptonic decay can be 
written as 
\begin{equation}
{\cal H}_{sl}= {4G_F\over {\sqrt 2}}V^L_{cb}[(\bar c_L \gamma_\mu b_L)+
\xi (\bar c_R \gamma_\mu b_R)](\bar \ell_L \gamma^\mu \nu_L)\,,
\end{equation}
where here we will treat $\xi$ as a {\it complex} parameter, 
$\xi=|\xi|e^{i\Delta}$, though CP violation will be ignored in the discussion 
that follows{\cite {prep}}. As we will see the additional phase degree of 
freedom will play 
a very important role in obtaining signals and constraints on RH currents. 
We recall that in the original Gronau and Wakaizumi 
scenario the leptonic current in $B$ decays was also RH{\cite {gro}} and 
neutrino masses were tuned to allow for the semileptonic decay process. 
How do we ascertain the allowed range of $\xi$? Again following Voloshin, 
the first place to obtain a constraint is inclusive semileptonic $b$ decay at 
the quark level. The most obvious observable is the inclusive decay partial  
width that can be written as 
\begin{equation}
\Gamma(b\to c\ell \nu)\sim |V^L_{cb}|^2 f(x)\eta_L \Bigg[1+|\xi|^2+2Re(\xi)
{g(x)\over {f(x)}} {\eta_R\over \eta_L}\Bigg]\,.
\end{equation}
For zero mass leptons, $f,g$ are the well-known kinematic phase space 
functions{\cite {vol,fuji,bsgme}} of the ratio $x=m_c/m_b \simeq 0.29$:
\begin{eqnarray}
f & = & (1-x^4)(1-8x^2+x^4)-24x^4\ln x \,, \\
g & = & -2x[(1-x^2)(1+10x^2+x^4)+12x^2(1+x^2)\ln x] \,, \nonumber
\end{eqnarray}
and for $x=0.29$ we find $f\simeq 0.542$ and $g\simeq -0.196$. For 
semileptonic decay to $\tau$'s, the corresponding phase space suppression 
factors can be decomposed as $f_\tau=(I_1+I_2)/2$ and $g_\tau=(I_1-I_2)/2$ 
in terms of the integrals
\begin{eqnarray}
I_1&=&\int_{y^2}^\Delta ds~Z\Bigg[\Delta(4s-y^2)+2\Delta\Sigma(1+2y^2/s)-(
\Sigma+2s)(2s+y^2)\Bigg]\,, \\
I_2&=&\int_{y^2}^\Delta ds~Z\Bigg[\Sigma(4s-y^2)+2\Delta\Sigma(1+2y^2/s)-(
\Delta+2s)(2s+y^2)\Bigg]\,, \nonumber
\end{eqnarray}
where $y=m_\tau/m_b$, $\Sigma=(1+x)^2$ and $\Delta=(1-x)^2$ with 
\begin{equation}
Z=\Bigg[1-{y^2\over {s}}\Bigg]^2\Bigg[(\Sigma-s)(\Delta-s)\Bigg]^{1/2}\,.
\end{equation}
Numerically, one finds $f_\tau \simeq 0.122$ and $g_\tau \simeq -0.0490$ for 
$x=0.29$ and $y\simeq 0.372$. Note that when RH currents are present the 
ratio of the $b$ 
semileptonic decay width to $\tau$'s to that for massless leptons becomes a 
weak function of $\xi$ with overall variations of order a few per cent; this 
dependence occurs due to a mismatch in the phase space ratios: 
$g_\tau/f_\tau \neq g/f$. This 
effect is most likely too small to be observed experimentally, however, but 
should be kept in mind.  

The parameters $\eta_{L,R}$ represent both perturbative and non-perturbative 
strong interaction corrections which also depend on $x$ as well as the lepton 
mass and the relevant strong interaction scale $\mu$. 
When needed in the numerical discussion below we will {\it assume} that the 
effects of all strong interaction corrections in the $b\to c$ 
semileptonic decay are 
at least approximately insensitive to the chirality of the charged current 
coupling, \ie, $\eta_L=\eta_R=\eta$ as was done Voloshin. (Certainly, an 
explicit calculation needs to be performed to verify this assumption.) 
To leading order in QCD for massless leptons and with $x=0.29$ one has the 
perturbative contributions 
$\eta=1-{2\alpha_s \over {3\pi}}(2.53)+{\cal O}(\alpha_s^2)$ whereas for 
$\tau$'s one obtains 
$\eta=1-{2\alpha_s \over {3\pi}}(2.11)+{\cal O}(\alpha_s^2)${\cite {hokim}}. 
The complete NLO 
expressions are not yet available in either case. Only the terms of order 
$\alpha_s^2\beta_0$, with $\beta_0$ being the one-loop QCD beta function, are 
known at present{\cite {luke}}, so for now we will truncate these corrections 
at this order but include them in our detailed numerical analysis below.

It is amusing to note that the existence of a RH coupling means that a 
measurement of the partial width $\Gamma$ yields not the true but an 
{\it effective} value of $V^L_{cb}$ from inclusive data when the result is 
interpreted in terms of the SM, \ie, 
\begin{equation}
|V^L_{cb}|^{inc}_{eff}=|V^L_{cb}|\cdot \Bigg[1+|\xi|^2+2Re(\xi){g\over {f}}
\Bigg]^{1/2}\,.
\end{equation}
under the assumption that $\eta_L=\eta_R$. This result will have important 
consequences for us below.

To obtain more information from this inclusive decay additional observables 
are required. 
Indeed, many authors have speculated about how one can experimentally extract 
information about potential RH couplings in inclusive semileptonic $b$ decay. 
Dittmar and Was{\cite {dw}} suggested examining simultaneously both the 
charged lepton and neutrino, \ie,  missing energy, spectra arising from $b$ 
semileptonic decay at the $Z$ peak. 
When one looks at the squared matrix element for this process in the 
free-quark limit, after all traces are performed, the sensitivity to $\xi$ 
becomes immediately transparent:
\begin{equation}
|{\cal M}|^2 \simeq p_\ell\cdot p_c p_\nu \cdot p_b+|\xi|^2 p_\ell \cdot p_b 
p_\nu \cdot p_c -m_bm_cp_\ell \cdot p_\nu Re(\xi)\,,
\end{equation}
with the $p_i$ labelling the corresponding particle four-momentum. The $\xi$ 
sensitivity is seen to be particularly enhanced due to the large value of the 
ratio $m_c/m_b \simeq 0.29$ with the phase of $\xi$ playing a very important 
role. (For completeness we note that one can find the full expression for the 
resulting unpolarized charged lepton spectra in the $Z$ rest 
frame at leading order is given by Fujikawa and Kawamoto{\cite {fuji}}. The 
corresponding neutrino spectrum can be trivially obtained through the 
interchange of the LH and RH couplings.) Interestingly, as mentioned earlier, 
L3{\cite {l3}} performed a simultaneous measurement of both the charged lepton 
and missing energy spectra in $b$ decay and excluded very large values of 
$\xi$, \ie, purely RH couplings, by more than $6\sigma$ and 
$\xi \simeq 1$, \ie, purely vector couplings, by more than $3\sigma$. They 
did not, however, attempt a fit to $\xi$ as the required sensitivity to 
$\xi << 1$ was not available once detector cuts and hadronic as well as other 
systematic uncertainties were taken into account. However, values of 
$|\xi|<< 1$ were certainly not excluded and we will attempt to further 
quantify these results below.

We now turn to each of the three experiments CLEO, ALEPH and L3 and survey 
the constraints that they are presently imposing on $\xi$ and what can be 
learned from comparable measurements at future $B$ factories even if 
relatively low integrated luminosities are available.

\subsection{CLEO}

In addition to inclusive semileptonic decay one may hope to obtain information 
on possible RH coupling through exclusive decay measurements due to the 
enriched nature of the accessible final states. In this regard 
CLEO{\cite {cleo}} has 
performed a detailed examination of both the $B\to D$ and $B\to D^*$ 
exclusive semileptonic modes. In the $B \to D$ case the impact of RH currents 
is well known to be rather minimal for massless leptons since the final state 
and the corresponding hadronic matrix element are rather simple. In 
this care, their only effect is to scale the anticipated partial rate by an 
overall factor, $|1+\xi|^2$, to which we will return below. A more complex 
and interesting pattern occurs for the $B\to D^*$ case. 

The CLEO analysis{\cite {cleo}} that examined the exclusive decay 
$B^0\to D^-*(\to D\pi)\ell \nu$ sought to extract form factor information and, 
in particular, to measure the forward-backward asymmetry of the charged 
lepton, $A_{FB}$, the average 
$D^*$ polarization, $\Gamma_L/\Gamma_T$,  as well as $V^L_{cb}$. The data 
sample of $\sim 780$ events employed in their analysis resulted from an initial 
set of $2.6\cdot 10^6$ $B\bar B$'s corresponding to an integrated 
luminosity of $\simeq 2.4 fb^{-1}$ at the $\Upsilon(4S)$.  
Following the general analysis as 
presented in Ref.{\cite {cleo,vold}}, one begins with an initially four-fold 
differential distribution but this is a bit unwieldy. Integration over 
two of the three decay 
angles (the others of which we will subsequently return to below) leads to the 
following double differential decay 
distribution for this process in the massless lepton limit:
\begin{equation}
{d^2\Gamma\over {dq^2dz}} \sim |V^L_{cb}|^2 Pq^2\Bigg[(1-z)^2|H_+|^2+(1+z)^2
|H_-|^2+2(1-z^2)|H_0|^2\Bigg]\,,
\end{equation}
where $P$ is the $D^*$ momentum in the $B$ frame, $q^2$ is the four-momentum 
transfer from the $B$ to the $D^*$ and $z=\cos \theta_\ell$ with $\theta_\ell$ 
being the decay angle of the $\ell$ in the virtual $W$ rest frame. $P$ is 
given by 
\begin{equation}
P={1\over {2M}} \Bigg[(M^2-m^2-q^2)^2-4m^2q^2\bigg]^{1/2}\,,
\end{equation}
and the 
helicity amplitudes $H_{\pm,0}$ are functions of $q^2$ which are generally 
expressed in terms of the conventional form factors $A_{1,2}$ and $V$ as 
\begin{eqnarray}
H_\pm(q^2) &=& (M+m)A_1(q^2)\mp{2MP\over {(M+m)}}V(q^2)\,, \nonumber \\
H_0(q^2) &=& [2m\sqrt {q^2}]^{-1} \Bigg[(M^2-m^2-q^2)(M+m)A_1(q^2)-{4M^2P^2
\over {(M+m)}}A_2(q^2)\Bigg]\,, 
\end{eqnarray}
where $M(m)$ is the mass of the $B(D^*)$. Meticulously following 
Neubert{\cite {adn}} one may use suggestive versions of the above form 
factors that have very well defined 
limits when Heavy Quark Effective Theory(HQET) becomes exact: 
\begin{eqnarray}
A_1(q^2) &=& {(M+m)\over {2\sqrt{Mm}}}\Bigg[1-{q^2\over {(M+m)^2}}\Bigg]h(w)
\,, \nonumber \\
A_2(q^2) &=& {(M+m)\over {2\sqrt{Mm}}}R_2(w)h(w)\,, \nonumber \\
V(q^2) &=& {(M+m)\over {2\sqrt{Mm}}}R_1(w)h(w)\,.
\end{eqnarray}
Here we define as usual $w=(M^2+m^2-q^2)/(2Mm)$. In the exact HQET limit both 
$R_{1,2}\to 1$ and $h(w)$ becomes the Isgur-Wise function so that the $R_i$ 
can be considered as representing small corrections in both $\alpha_s$ and 
$1/m$ to the case of pure leading order HQET. Generically, 
$h$ has a linear form, $h(w)=h(1)[1-\rho^2(w-1)]$ although other structures are 
possible. 

While the forward-backward asymmetry can be obtained by integration of the 
expressions above, the ratio $\Gamma_L/ \Gamma_T$ can be determined from the 
decay angular distribution of the $D$ in the $D^*$ frame when $D^*\to D\pi$ 
($\cos \theta_V$, in the notation of Ref.{\cite {cleo}}). Following 
Ref.{\cite {cleo,vold}} we can write the relevant double-differential 
distribution in this case as 
\begin{equation}
{d^2\Gamma\over {dq^2d\cos \theta_V}} \sim Pq^2\Bigg[(|H_+|^2+|H_-|^2)(1-
\cos^2 \theta_V)+2|H_0|^2 \cos^2\theta_V \Bigg]\,.
\end{equation}
$\Gamma_L/ \Gamma_T$ essentially probes the relative weights of the $H_0$ and 
$H_\pm$ helicity amplitudes as we will see shortly.

So far this discussion has been quite general. To include the effects of 
$\xi \neq 0$ in comparison to SM expectations we simply make the replacements 
$V\to V(1+\xi)$ and $A_{1,2}\to A_{1,2}(1-\xi)$ in the expressions for the 
helicity amplitudes 
above and recall that $\xi$ is complex. This follows directly from the 
rescaling of the LH and RH current amplitudes as seen in Eq.(1). Once 
particular expressions for $R_{1,2}$ and $h$ are assumed we may 
directly calculate $A_{FB}$, $\Gamma_L/\Gamma_T$, as well as the total decay 
rate, which then gives us $V_{cb}^{L~exc}(D^*)$. We obtain 
\begin{eqnarray}
A_{FB}&=&{\int dq^2[\int_0^{z_0}-\int_{-z_0}^0]dz ~{d^2\Gamma\over 
{dq^2dz}}
\over {\int dq^2\int_{-z_0}^{z_0}dz ~{d^2\Gamma\over {dq^2dz}}}}\,, 
\nonumber \\
{\Gamma_L\over {\Gamma_T}}&=&{\int dq^2 ~2z_0(1-z_0^2/3)Pq^2 H_0^2\over {\int 
dq^2 ~z_0(1+z_0^2/3)Pq^2 (H_+^2+H_-^2)}}\,,
\end{eqnarray}
where $z_0(q^2)$ expresses a potential minimum lepton momentum cut used to 
identify the event:
\begin{equation}
z_0=min\Bigg[1, -{4Mp_\ell^{cut}-M^2-q^2-m^2\over {2PM}}\Bigg]\,.
\end{equation}
CLEO, for example, employs a typical lepton momentum cut of $\simeq 1$ GeV. 
These expressions can be re-written to clearly display their $\xi$ dependence 
as 
\begin{eqnarray}
A_{FB} &=& {(1-|\xi|^2)C\over {(1+|\xi|^2)A-2BRe(\xi)}}\,, \nonumber \\
{\Gamma_L\over {\Gamma_T}} &=& {4\over {3}} {[1+|\xi|^2-2Re(\xi)]D\over 
{(1+|\xi|^2)E+2FRe(\xi)}}\,.
\end{eqnarray}
Experimentally, CLEO{\cite {cleo}} finds $A_{FB}=0.197\pm 0.037$ and 
$\Gamma_L/\Gamma_T=1.55\pm 0.29$, which are of course both consistent with 
SM/HQET expectations. Here $A-F$ are a simple set of numbers which result 
from performing the double integration over the 
relevant kinematics. For a fixed set of $R_{1,2}$ and $h$, the values of $A-F$ 
are completely determined subject to experimental cuts, and these results can 
be combined to constrain $\xi$. In addition, from the expression for the 
overall partial width we also obtain 
\begin{equation}
|V^L_{cb}|^{exc}_{eff}(D^*)=|V^L_{cb}|\cdot \Bigg[1+|\xi|^2-2Re(\xi){B\over {A}}
\Bigg]^{1/2}\,,
\end{equation}
when the value is again interpreted in the SM; this result should then be 
compared with Eq.(6). 
Note that since one finds that $-B/A \neq g/f$, the apparent values of 
$V^L_{cb}$ extracted from exclusive $B\to D^*$ and inclusive measurements 
will be {\it different} when $Re(\xi)\neq 0$. 
Demanding that the {\it true} $V^L_{cb}$ take on the same value 
in both cases imposes an extra constraint on $\xi$. In order to employ 
this additional constraint we use the specific numerical results as provided 
in the recent review of both inclusive and exclusive semileptonic decay 
data by Buras{\cite {buras}} to obtain 
$V_{cb}^{L~exc}(D^*)/V_{cb}^{L~inc}=0.967\pm 0.105$. This value is completely 
consistent with unity, as anticipated, but will still provides an additional 
requirement on $\xi$. A similar situation, as mentioned above, occurs in the 
case of $B\to D$ semileptonic decays where we now would find simply 
\begin{equation}
|V^L_{cb}|^{exc}_{eff}(D)=|V^L_{cb}|\cdot \Bigg[1+|\xi|^2+2Re(\xi)
\Bigg]^{1/2}\,,
\end{equation}
which should be compared with that from the $D^*$ mode above. 
Given the present experimental situation{\cite {cleo}} adding this additional 
constraint will 
not influence the results of the fit obtained below. However, future 
measurements may make this an important input into analyses of RH currents. 

Our procedure is the following: for a fixed set of $R_{1,2}(w)$ and $h(w)$ 
we calculate the integrals $A-F$ and then perform a simultaneous 
$\chi^2$ fit to 
the CLEO results on $A_{FB}$ and $\Gamma_L/\Gamma_T$ as well as to the ratio 
$V_{cb}^{L~exc}(D^*)/V_{cb}^{L~inc}$ treating $|\xi|$ and 
$c_\Delta=\cos \Delta$ as 
free parameters (recall, $\Delta$ is the phase of $\xi$). Possible 
correlations are ignored. We then choose 
another set of $R_{1,2}$ and $h$ and repeat the process. Each repetition will 
thus generate a $95\%$ CL allowed region in the $c_\Delta-|\xi|$ plane. To 
be specific we employ forms of $R_{1,2}(w)$ and $h(w)$ suggested by 
Neubert{\cite {adn}} and by Close and Wambach(CW){\cite {cw}} as well as 
several other sets suggested by the first paper in Ref.{\cite {cleo}}. As a 
typical 
example, with $R_1^{CW}=1.15[1-0.06(w-1)]$, $R_2^{CW}=0.91[1+0.04(w-1)]$ 
and $\rho^2=0.91$ we obtain $A\simeq 0.116$, $B\simeq 0.105$, $C\simeq 0.024$, 
$D\simeq 0.060$, $E\simeq 0.056$, and $F\simeq 0.044$. These values do indeed 
reproduce the well known SM expectations{\cite {adn}} in the $\xi \to 0$ limit.

\vspace*{-0.5cm}
\nn
\begin{figure}[htbp]
\centerline{
\psfig{figure=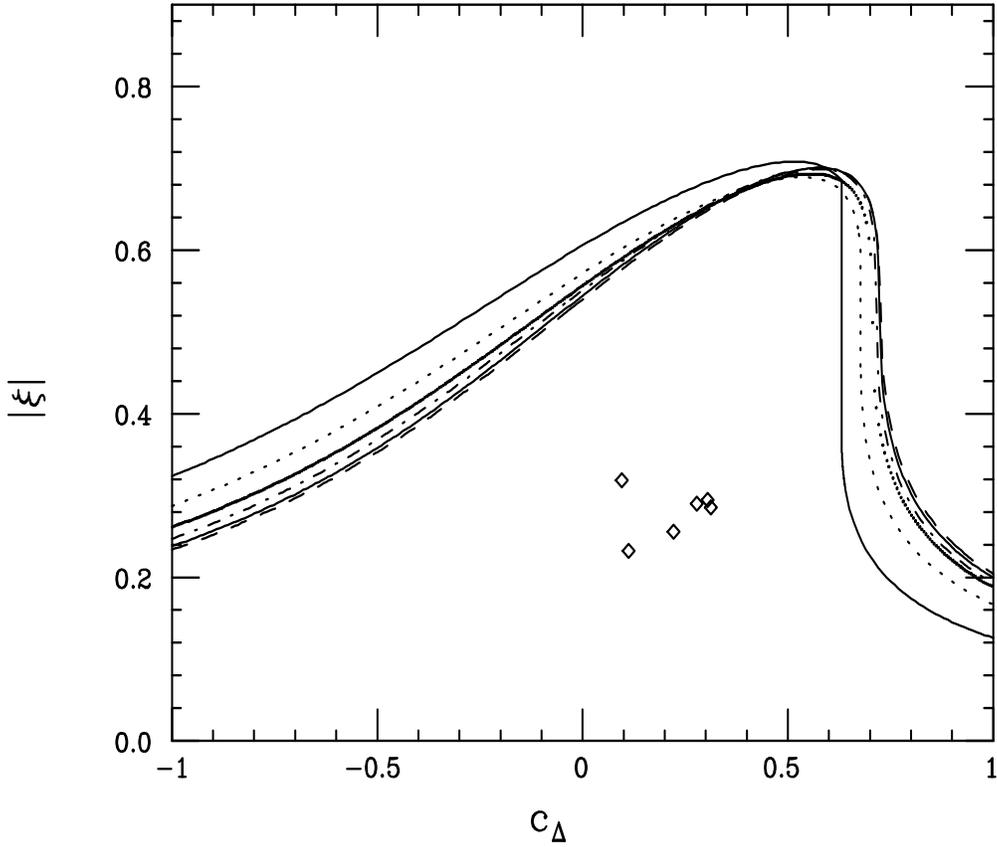,height=14cm,width=16cm,angle=-90}}
\vspace*{-1cm}
\caption{$95\%$ CL allowed region (below the curves) in the $|\xi|-c_\Delta$ 
plane obtained from a fit to CLEO data as well as the experimental value of 
the ratio $V_{cb}^{L~exc}(D^*)/V_{cb}^{L~inc}$. Each of the six curves 
corresponds 
to a unique choice of $R_{1,2}$ and $h$. The SM limit lies along the 
horizontal axis at $|\xi|=0$. The 
locations of the six $\chi^2$ minima are also shown for completeness and are 
seen to be reasonably clustered.}
\label{hqet}
\end{figure}
\vspace*{0.4mm}

The results of this fit are shown in Fig.~\ref{hqet} which displays the 
$95\%$ CL upper bound on $|\xi|$ as a function of $c_\Delta$ for several 
different choices of $R_{1,2}$ and $h$. The most important features of these 
results to notice are: ($i$) 
the bounds we obtain are not very sensitive to the exact choice of these HQET 
functions and ($ii$) the constraints on $|\xi|$ are strongest when $\xi$ is 
real. We note that Voloshin's preferred range of values of $\xi=0.14\pm 0.18$ 
lie mostly inside the allowed region. It is clear that at the moment the 
existing constraints on $\xi$ are quite poor and that values of $|\xi|$ of 
order 0.25 are certainly allowed by current data. We note that for the six 
sets of HQET functions used in this analysis the resulting best fit values 
for $\xi$ are reasonably clustered and indicate a magnitude 
$\simeq 0.20-0.35$ and a sizeable phase. With the far larger data sets soon 
to be available from 
$B$ factories it is quite important for this analysis to be be revisited and 
refined in the not too distant future. We note that a somewhat smaller allowed 
region results if the unpublished results from CLEO that now include the 
charged $B$ decay modes are used{\cite {unp1}}.

\vspace*{-0.5cm}
\nn
\begin{figure}[htbp]
\centerline{
\psfig{figure=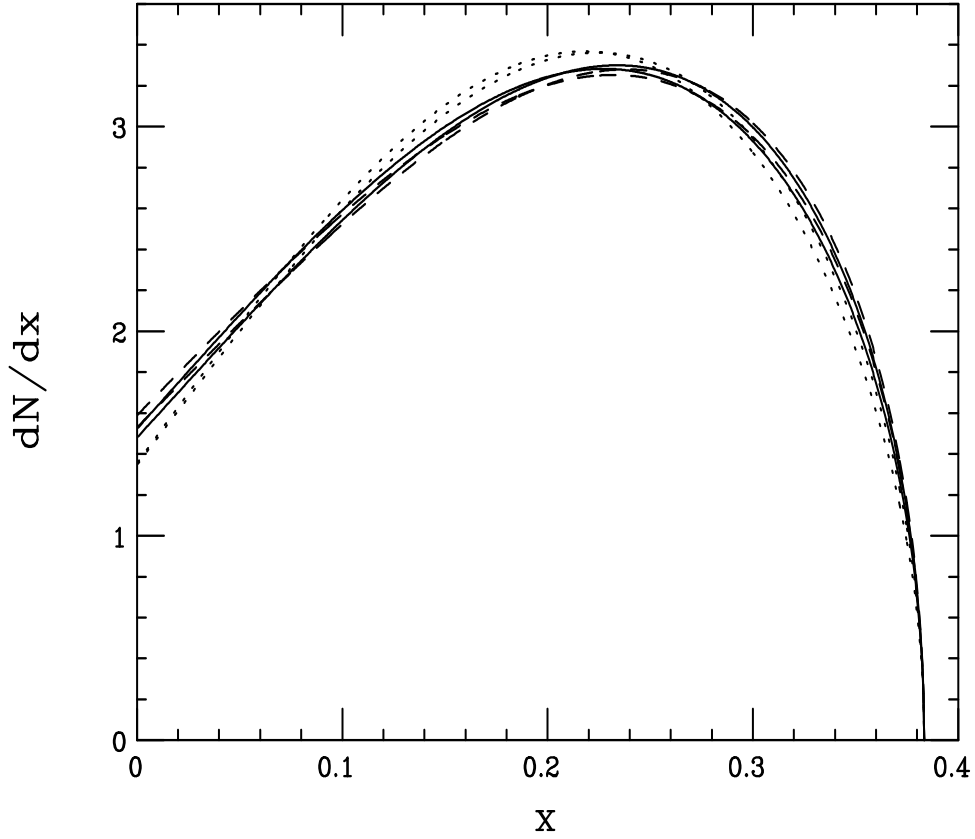,height=14cm,width=16cm,angle=-90}}
\vspace*{-1cm}
\caption{Normalized $q^2$ distributions for the process $B\to D^*\ell \nu$. 
Here $x=q^2/M^2$ and the curves correspond to the SM(solid) and 
$\lambda=0.5(-0.5)$(dotted and dashed, respectively). A possible $p_t$ cut 
on the charged lepton momenta has been ignored. Results are shown for both the 
Neubert as well as the Close and Wambach HQET functions corresponding to the 
pair of curves for each case.}
\label{q2dis}
\end{figure}
\vspace*{0.4mm}

One might ask if there are other observables associated with this exclusive 
decay that could allow for some additional sensitivity to RH current 
interactions{\cite {vold}}. To this end we 
briefly examine both the $q^2$ and $\chi$ distributions which can be measured 
using the recoil momentum of the $D^*$ and identifying 
the angle between the $W$ and $D^*$ 
event planes, respectively. Once integrated over all other variables, 
the deviations in both these distributions from the SM expectations are found 
to be totally 
controlled by the value of the ratio $\lambda=2|\xi|c_\Delta/(1+|\xi|^2)$. 
The form of the 
$q^2$ distribution can be obtained immediately from the double differential 
expression above. Fig.~\ref{q2dis}, 
where we have used the HQET functions, $R_i$, of Neubert{\cite {adn}} and 
those of 
Close and Wambach{\cite {cw}}, shows that the normalized distribution is 
only very weakly dependent on the existence of RH currents. Specifically, we 
see a direct comparison of the SM distribution, $\lambda=0$, with that 
expected for the cases of $\lambda=\pm 0.5$. From the figure it appears 
unlikely that the shape of the $q^2$ distribution will yield any useful 
information on RH currents unless very high precision data is obtainable.
Note there is little difference between the curves generated with the 
two different sets of HQET functions. 

In the case of the normalized $\chi$ distribution, the shape is controlled by 
a single parameter {\it if} all the other variables have been completely 
integrated over, \ie, 
\begin{equation}
{dN\over {d\chi}}={1\over {\pi}} ( 1-\Omega \cos 2\chi) \,, 
\end{equation}
where 
\begin{equation}
\Omega={\int dq^2 ~Pq^2 Re(H_+^*H_-)\over {\int dq^2 ~Pq^2 
(H_+^2+H_-^2+H_0^2)}}\,,
\end{equation}
which we may rewrite to show the $\lambda$ dependence explicitly as
\begin{equation}
\Omega=-{(T_2+T_1\lambda)\over {(2T_1+T_3)+(2T_2-T_3)\lambda}}\,,
\end{equation}
with the $T_i$ being a set of kinematic integrals. 
In the SM one finds that $\Omega \simeq 0.175(0.192)$ using Neubert(CW) HQET 
functions. Note that if instead only the 
even(odd) values of $\cos \theta_V$ are integrated over, the 
normalized $\chi$ distribution picks up an additional term of the form
\begin{equation}
{dN\over {d\chi}}\to {dN\over {d\chi}}\mp {3\over {8}}\Sigma \cos \chi \,, 
\end{equation}
where 
\begin{equation}
\Sigma={\int dq^2 ~Pq^2 Re~H_0^*(H_+-H_-)\over {\int dq^2 ~Pq^2 
(H_+^2+H_-^2+H_0^2)}}\,.
\end{equation}
For this variable the $\xi$ and $c_\Delta$ dependencies become are somewhat 
more complex and cannot be expressed simply through the parameter $\lambda$; 
$\Sigma$ is expressible as 
\begin{equation}
\Sigma={T_4{\mbox {$(1-|\xi|^2)$}\over \mbox {$(1+|\xi|^2)$}}\over {(2T_1
+T_3)+(2T_2-T_3)\lambda}}\,,
\end{equation}
with $T_4$ being another kinematic integral. In the SM one finds that 
$\Sigma \simeq -0.25(-0.22)$ for Neubert(CW) HQET functions.

\vspace*{-0.5cm}
\nn
\begin{figure}[htbp]
\centerline{
\psfig{figure=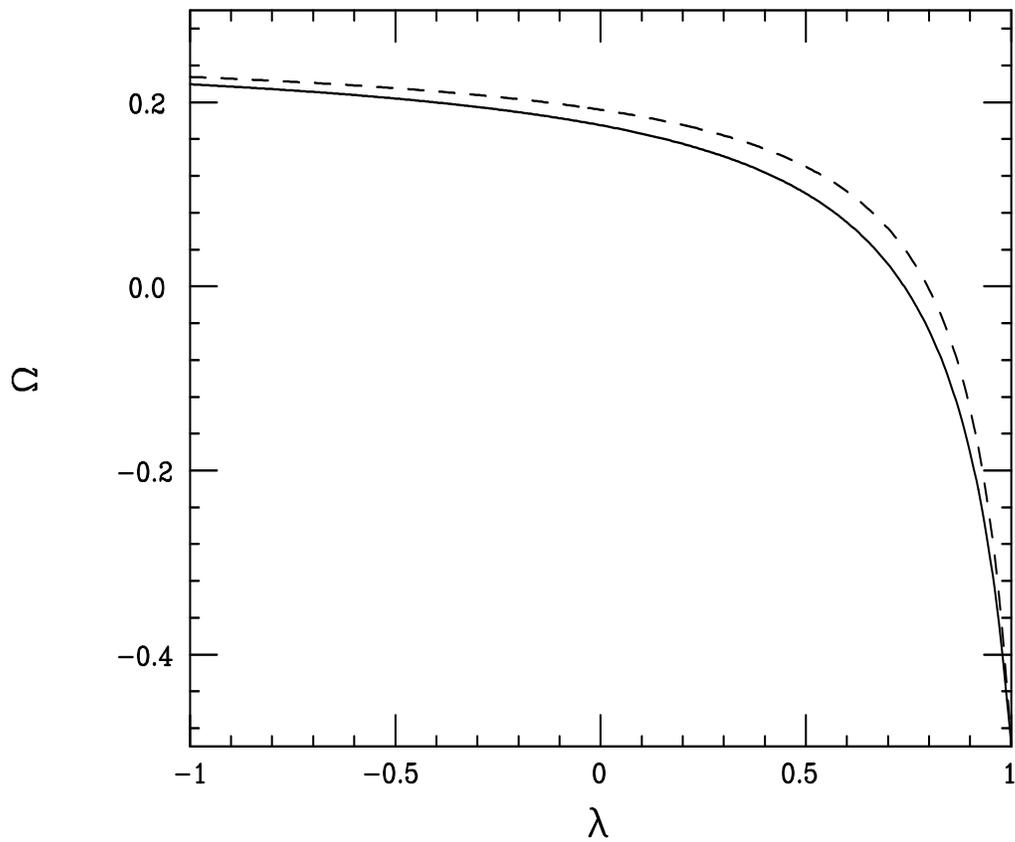,height=14cm,width=16cm,angle=-90}}
\vspace*{-1cm}
\caption{Value of the $\Omega$ parameter which controls the shape of the 
$\chi$ distribution as a function of $\lambda$. The results are shown for both 
the Neubert(solid) as well as the Close and Wambach(dashed) HQET functions 
which are seen to yield quite similar results.}
\label{omega}
\end{figure}
\vspace*{0.4mm}

Fig.\ref{omega} shows that $\Omega$ is quite sensitive to positive values of 
$\lambda$ so that one may hope to get a reasonable sensitivity to RH 
interactions if $\Omega$ could be precisely measured. Present data from CLEO 
is found to be consistent{\cite {cleo}} with the expectations of the 
SM for $\Omega$ but the statistics are still rather poor. To get an idea of the 
potential sensitivity we have performed a straightforward two 
parameter (normalization and $\Omega$) fit to 
the existing binned data as presented in Ref.{\cite {cleo}}. To obtained 
improved statistics in this first fit we have combined the data 
in both the $\cos \theta_V >0$ 
and $\cos \theta_V <0$ regions. Unfortunately, the resulting the distribution 
of the data shows little sensitivity to $\Omega$. 
After background subtraction this fit yields $\Omega=0.126\pm 0.120$ at 
$95\%$ CL which is certainly consistent with the SM. This constraint 
subsequently implies that $\lambda$ lies in 
the $95\%$ CL range $-3.3(-2.8)\leq \lambda \leq 0.71(0.75)$ for Neubert(CW) 
HQET functions using the results in Fig.\ref{omega}. As one would 
expect from the low sensitivity to negative $\lambda$, our bound in this 
case is rather poor. 

\begin{figure}[htbp]
\centerline{
\psfig{figure=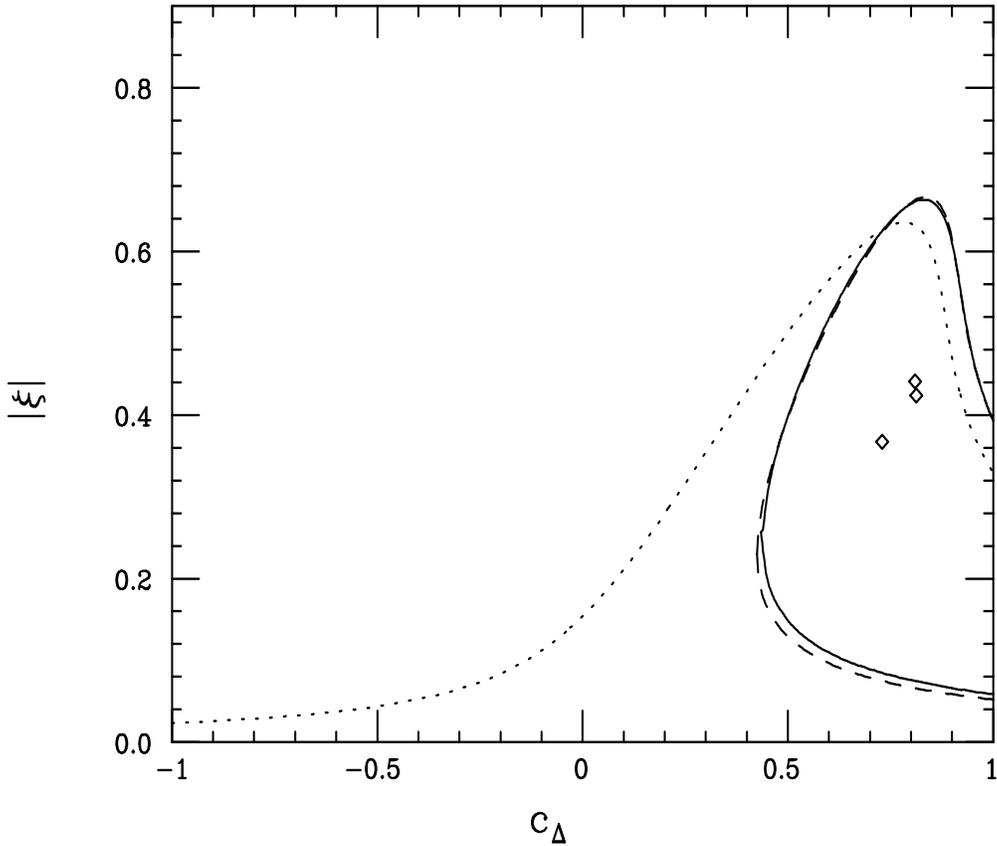,height=14cm,width=16cm,angle=-90}}
\vspace*{-1cm}
\caption{$95\%$ CL fit to the shape of the CLEO $\chi$ distribution assuming 
Neubert(dotted), CW(dashed) or ISGW(solid) HQET functions. The allowed region 
is either below the dotted line or within the dashed or solid enclosure. 
As before the diamonds locate the $\chi^2$ minima for the three sets of $R_i$.}
\label{fun}
\end{figure}

A second more hopeful possibility is to fit the shape of the $\chi$ 
distribution 
for {\it both} $\Omega$ and $\Sigma$ by treating the $\cos \theta_V>0$ and 
$\cos \theta_V<0$ regions independently; here there is a loss of statistics 
but a dramatic increase in sensitivity to RH couplings. 
Following the same analysis as above we arrive at the results presented in 
Fig.\ref{fun} for the allowed region in the $c_\Delta-|\xi|$ plane for 
Neubert, CW as well as Isgur and Wise{\cite {isgw}} HQET functions. Note 
that the allowed region resulting from this fit is somewhat sensitive to the 
HQET $R_i$ choice, quite unlike the other observables that we have examined 
up to this point. Although this result seems to support the possibility 
that RH currents may indeed be present one must be hesitant to form such a 
hasty conclusion without further analysis. First, the only believable fit of 
this kind must be performed by the CLEO Collaboration and we note the 
apparent strong sensitivity of our result to the choice of the $R_i$ 
HQET functions. However, it is certainly most clear that our understanding of 
potential RH currents in $b$ decay would very much profit from higher 
precision measurements of the $\chi$ distribution. This seems possible during 
the first year of $\Upsilon(4S)$ running of BABAR and BELLE since the CLEO 
data sample used in this analysis corresponded to only 2.6 million $B\bar B$ 
pairs. We note in passing that using the still unpublished CLEO data from 
the charged $B$ decay mode{\cite {unp1}} already strengthens the case for 
right-handed couplings based on the fit to the $\chi$ distribution. 

Another question one might ask is what the allowed range for the parameters 
$\Omega$ and $\Sigma$ are given the CLEO constraints we have extracted from 
the earlier fit. To obtain such results we need to scan the $|\xi|-c_\Delta$ 
region below the envelope of curves shown in 
Fig.1 to get the extrema. We find $0.053\leq \Omega \leq 0.207$ and 
$-0.345 \leq \Sigma \leq -0.115$ for the Neubert HQET functions; 
correspondingly, for the CW HQET functions we obtain 
$0.089\leq \Omega \leq 0.218$ and $-0.310 \leq \Sigma \leq -0.106$. 

\vspace*{-0.5cm}
\nn
\vspace*{0.4mm}

As a final note, if the $\tau$ polarization in the decay $B\to D^*\tau \nu$ 
can be measured, Wakaizumi has shown{\cite {wak}} that it provides yet another 
quantity which is fairly sensitive to $\xi \neq 0$.  This decay mode will thus 
yield even more observables which can be used to probe for $b\to c$ RH 
currents due to the addition of finite mass terms associated with the $\tau$. 
Of course for this mode there is a loss in statistics due to the additional 
phase space suppression due to the $\tau$ mass as well as the associated $\tau$ 
reconstruction efficiency to be dealt with. An analysis of these prospects 
is, however, beyond the scope of the present work{\cite {prep}}.

\subsection{ALEPH}

Unfortunately, other data cannot at present improve significantly upon the
CLEO bounds without further employing some rather strong assumptions. 
For example, in principle the low $\Lambda_b$ polarization observed in $Z$ 
decay{\cite {lambda}} by ALEPH can be used to obtain such a constraint. 
We recall that a $b$ quark produced at the $Z$ in the SM is highly 
polarized, \ie, $P=-0.935$ and radiative effects have been shown to reduce 
this value only slightly{\cite {slight}}. During the hadronization process 
some of the memory of the original $b$ polarization is lost but it had been 
anticipated that in the $b\to \Lambda_b$ process a large part of the original 
polarization would be kept{\cite {keep}}. Falk and Peskin{\cite {keep}} 
estimated on the basis of HQET that the resulting $\Lambda_b$ polarization 
would be $P=-(0.69\pm 0.06)$. 

The ALEPH analysis is based on $\sim 3\cdot 10^6$ hadronic $Z$ decays which 
yielded a sample of $462\pm 31$ $\Lambda_b$ candidates. 
The method used by ALEPH to extract the value of $P$ for the $\Lambda_b$ 
was first suggested by Bonvicini and 
Randall{\cite {rand}} who noted that the ratio of the average values of the 
neutrino and lepton energies in semileptonic $B$ decay, $y=<E_\ell>/<E_\nu>$, 
was particularly sensitive to the polarization of the $b$ quark. This 
variable, being an energy ratio, 
is quite insensitive to $b$ fragmentation, detector acceptance and 
reconstruction effects as well as the uncertainties in the ratio $m_c/m_b$. 
We note that the direct comparison of the average 
charged lepton and neutrino energies from $b$ quark decays with theoretical 
expectations, as was done by L3{\cite{l3}}, was found to lead to substantial 
fragmentation uncertainties although values of $\xi$ of order unity were 
clearly excluded; we will return to the L3 data below. 
It has also been found that 
$\alpha_s$ and $1/m_b^2$ corrections{\cite {corr}} to the parton level 
expectations for $y$ are quite small (and, hence, will subsequently 
be ignored). Of course these results have only been explicitly demonstrated in 
the case of purely LH couplings. In our analysis we will make the reasonable 
assumption that they remain 
true when both LH and RH couplings are present. 

The averages of $E_\ell$ and 
$E_\nu$ can be calculated directly from the decay at rest spectra through 
the boost relations $E_\ell=\gamma(E_\ell^*+\beta p_L^*)$, ~\etc.,  with 
$\beta \simeq 1$ and $p_L^*$ being the lepton's momentum in the boost 
direction. In order to remove selection 
cut and energy reconstruction errors which produce a bias in $y$, ALEPH 
instead determined the ratio of ratios 
$R_y=y_{data}/y_{MC}(0)$ where $y_{MC}(0)$ is 
the $y$ value obtained from a Monte Carlo simulation employing the SM in the 
limit of zero polarization. The value of $R_y$ was then compared with the 
Monte Carlo-corrected SM theory prediction to extract the value of $P$. 
What ALEPH found was $R_y=1.10\pm 0.13$ (including systematic errors in 
quadrature)that then yielded the intriguingly small value 
$P=-0.23^{+0.26}_{-0.23}$, which is significantly smaller in magnitude, by 
$\simeq 2\sigma$, than the expectations of Falk and Peskin. 

\begin{figure}[htbp]
\centerline{
\psfig{figure=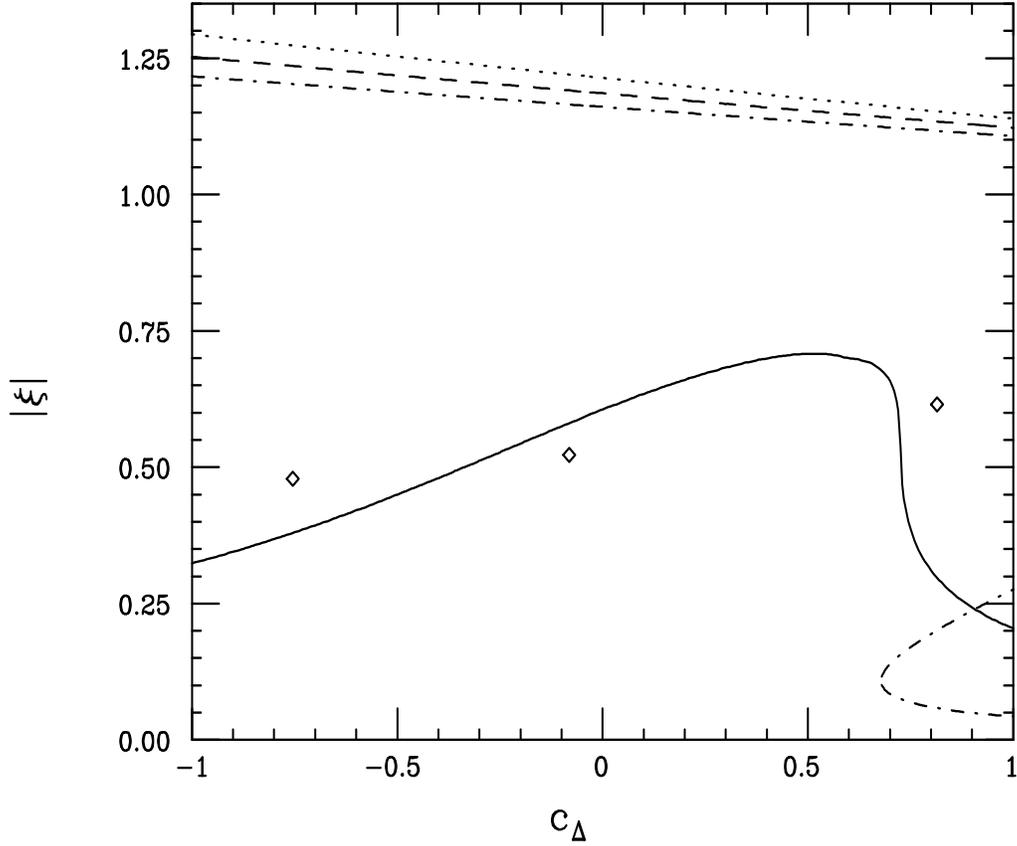,height=14cm,width=16cm,angle=-90}}
\vspace*{-1cm}
\caption{Comparison of the envelope of the $95\%$ CL allowed regions obtained 
with CLEO data
(solid curve) with those obtainable from the ALEPH $\Lambda_b$ polarization 
results assuming $P=-(0.69\pm 0.06)$ corresponding to the dotted, dashed and 
dash dotted curves. The region below the slightly tilted horizontal curves 
and outside the `nose' on the lower right-hand side for the case of 
$P=-0.75$ are allowed. The location of the corresponding $\chi^2$ minima for 
$P$=$-0.63$, $-0.69$, and $-0.75$, respectively, are also displayed from 
right to left.}
\label{lambda}
\end{figure}

To investigate the double ratio $R_y$ in the case when RH currents are 
present, we must return to the 
normalized double-differential charged lepton decay distribution. In the $b$ 
rest frame to leading order and neglecting the lepton mass we find 
\begin{equation}
{dN\over {dzd\cos \theta}}=\Bigg[{R(x,z)+P\cos \theta ~Q(x,z)
\over {(1+|\xi|^2)f(x)+2Re(\xi)g(x)}}\Bigg]\,,
\end{equation}
where $z=2E_\ell/m_b$ and $\theta$ is the angle between the $b$ and $\ell$ 
momenta with $f(x)$ and $g(x)$ given above. Explicitly, we find that 
$R=R_{LL}+R_{RR}|\xi|^2+2Re(\xi)R_{RL}$ and $Q=Q_{LL}+Q_{RR}|\xi|^2+
2Re(\xi)Q_{RL}$ with 
\begin{eqnarray}
R_{LL}&=&{z^2(1-x^2-z)^2\over {(1-z)^3}}\Bigg[(1-z)(3-2z+x^2)+2x^2\Bigg]
\,, \nonumber \\
R_{RR}&=&{6z^2(1-x^2-z)^2\over {(1-z)}}\,, \\
R_{LR}&=&-{6xz^2(1-x^2-z)^2\over {2(1-z)^2}}\,, \nonumber 
\end{eqnarray}
and 
\begin{eqnarray}
Q_{LL}&=&{z^2(1-x^2-z)^2\over {(1-z)^3}}\Bigg[(1-z)(1-2z+x^2)-2x^2\Bigg]
\,, \nonumber \\
Q_{RR}&=&{6z^2(1-x^2-z)^2\over {(1-z)}}\,, \\
Q_{LR}&=&-{6xz^2(1-x^2-z)^2\over {2(1-z)^2}}\,. \nonumber 
\end{eqnarray}
These results confirm those obtained by Tsai{\cite {tsai}} long ago in a 
different form and context. The corresponding expressions for the neutrino 
spectrum can be obtained from the explicit relations above by interchanging 
the role of the left- and right-handed labels. 
Using these results we can calculate $y$ following Bonvicini and 
Randall{\cite {rand}}, rescale this value by the SM result assuming $P=0$, 
and include the Monte Carlo corrections of ALEPH. Given an assumed value for 
$P$ we can then fit to the ALEPH data to obtain an allowed region in the 
$|\xi|-c_\Delta$ plane. The result of this analysis is shown in 
Fig.~\ref{lambda} and is compared to the CLEO allowed region obtained above 
{\it assuming} the estimate of the polarization retention of Falk and Peskin, 
$P=-(0.69\pm 0.06)$, is correct. Here we see that at 
the $95\%$ CL almost the entire plane is allowed except for a possible 
small region on the lower right which only appears in the case of $P=-0.75$. 
As $P$ increases in magnitude, we note that 
the allowed parameter space region shrinks somewhat in size. We also see from 
the figure that the location of the best fit is quite sensitive to the 
assumed true value of the polarization. We note that if 
there are additional dynamical mechanisms{\cite {sal}} which could lead to a 
further reduction in the expected value of $P$ in the SM {\it and} they could 
be reliably trusted quantitatively, then the limits we would obtain on RH 
$b\to c$ couplings might be improved. In the future if the central value 
obtained by ALEPH 
was verified and the errors were reduced by a factor of two the size of the 
allowed region would shrink substantially and form a band approximately 
$\delta |\xi|\simeq \pm 0.25$ wide on 
either side of the best fit points shown in the Figure. 

\vspace*{-0.5cm}
\nn
\vspace*{0.4mm}

If the apparent low value of the polarization as measured by ALEPH is verified 
by future experiments then there are only two conclusions. Within the SM 
framework there must be some new source of depolarization and indeed 
$P\simeq -0.23$. Alternatively, right-handed 
currents are present and the true value of $P$ is closer to the HQET 
expectations of $P\simeq -0.69$ but {\it appears} low when interpreted in terms 
of the SM. As discussed above, a reduction in the ALEPH error by a factor of 
two, assuming the 
same central value, would clearly define a small allowed region when combined 
with the results from CLEO.  
Unfortunately, a measurement of $R_y$ {\it alone}, no matter how precise, 
will be able to eliminate the possibility of some exotic depolarization 
mechanism and allow us to conclude that RH couplings exist. However, an 
analysis of the higher $y$ moments or other possible distributions may be most 
helpful as suggested by Diaconu \etal {\cite {rand}} in an important 
paper. For simplicity we 
first consider only the ratio of the second moments of the decay 
distributions here. (We have not examined moments higher than second.) 
Within the SM if $x=0.29$ and $P=-0.23$ we can uniquely 
predict the value of the quantity $R_{2y}=y_2/y_2(0)=1.181$,  where 
$y_2=<E^2_\ell>/<E^2_\nu>$, although $\alpha_s$ and $1/m_b^2$ corrections are 
somewhat larger here. In the case where RH currents are present and $P=-0.69$, 
we can invert the $R_y$ relation and find $|\xi|$ as a function of $c_\Delta$ 
and then calculate the corresponding value of $R_{2y}$. We find in this case 
that for the central value $R_y=1.10$ one obtains 
$1.198 \leq R_{2y} \leq 1.227$ apart from the above corrections; note that 
this range does not overlap with the SM expectation but the separation between 
the two is quite small. It would thus 
appear that simultaneous very high precision measurements of both 
$R_y$ and $R_{2y}$, as well as possible higher moments, 
are required in order to be able to resolve the ambiguity and determine if RH 
currents are indeed present in semileptonic $b$ decays. Given the current and 
anticipated sizes of the errors a determination of at least these first two 
moments alone will not necessarily prove useful. 

As a second possibility we note that 
Diaconu \etal ~also suggest a number of other variables which can be used 
to probe the $\Lambda_b$ polarization. One of these is the difference in the 
charged lepton and neutrino rapidities, $\Delta \eta=\eta_\ell-\eta_\nu$, where 
these rapidities are measured with respect to the boost direction. This 
quantity is directly proportional to the polarization and, being a rapidity 
difference, is fortunately insensitive to fragmentation uncertainties. 
We find that with RH currents contributing one obtains 
\begin{equation}
\Delta \eta=P\int_{-1}^{1}d\cos \theta ~\eta (1-|\xi|^2){\int ~dz (Q_{LL}-
Q_{RR})\over {(1+|\xi|^2)f(x)+2Re(\xi)g(x)}}\,,
\end{equation}
where $\eta={1\over {2}} log{(1+\cos \theta)\over {(1-\cos \theta)}}$. 
Numerically we confirm the SM result and more generally obtain 
\begin{equation}
\Delta \eta \simeq {-0.632P(1-|\xi|^2)\over {(1+|\xi|^2)+2|\xi| c_\Delta(-0.362)
}}\,,
\end{equation}
so that in the SM for $P=-0.69(-0.23)$ we would 
obtain $\Delta \eta=0.436(0.145)$. 
In the case of RH currents, repeating the above procedure to find $|\xi|$ as a 
function of $c_\Delta$ from the data on $R_y$ we are led to the prediction 
that $\Delta \eta=0.238-0.257$, assuming that $P=-0.69$, which is quite 
different from either the SM expectation with a low value of $P$ or the HQET 
SM prediction. Again it appears that the RH current and exotic depolarization 
mechanism possibilities may be separable using precision measurements. 
However in this case we note that the required level of precision for 
this variable is far less that that for $R_{2y}$ giving us some hope that 
such a separation may indeed be possible at future $B$ factories{\cite {unp2}.

\subsection{L3}

Following the same approach as above one might attempt to further quantify the 
L3{\cite {l3}} constraints on $\xi$ by constructing the $y$ values using the 
results presented in their Table 2 and including some corrections associated 
with their Monte Carlo. This would then be similar to the ALEPH analysis but 
now one is actually probing 
the initial $b$ quark polarization about which there is far less uncertainty. 
Of course, in principle, only L3 can perform this procedure but our 
rudimentary study  will provide an indication for the location and size of the 
allowed region associated with their data.  
If we simply double their errors but then ignore both the 
$\alpha_s$ and $1/m_b^2$ corrections as well as fragmentation and 
energy scale uncertainties and neglect any correlations, we can 
obtain an estimate for the associated allowed region in the 
$c_\Delta-|\xi|$ plane. This most likely substantially underestimates the 
present experimental and theoretical 
uncertainties. Here we also need to input the parton-level polarization, 
$P=-0.935$. The results 
of these questionable considerations are shown in Fig.~\ref{l3stuff} and are 
compared to the 
CLEO analysis constraints.  From this figure we see that the crude estimate 
of the L3 constraints and those obtained above from CLEO are not in conflict 
and even tend to prefer similar regions of the parameter space. The sizes of 
the two allowed regions are rather comparable and substantially overlap.  
It is also clear from the figure that the L3 data certainly excludes both 
a $(V+A)\times (V-A)$ as well as a $V\times (V-A)$ interaction as several 
$\sigma$ as claimed.
Before we can draw any stronger conclusions, however, this analysis needs to be 
repeated by L3 themselves with the additional $\alpha_s$ and $1/m_b^2$ 
corrections included. We can conclude that future spectra determinations from 
inclusive decays will indeed be useful in probing for RH currents provided 
high statistics are available and systematic experimental uncertainties are 
under control. Since the L3 analysis is based on a sample of only $10^6$ 
$Z$'s it is clear that a higher statistical study can be performed. 

\vspace*{-0.5cm}
\nn
\begin{figure}[htbp]
\centerline{
\psfig{figure=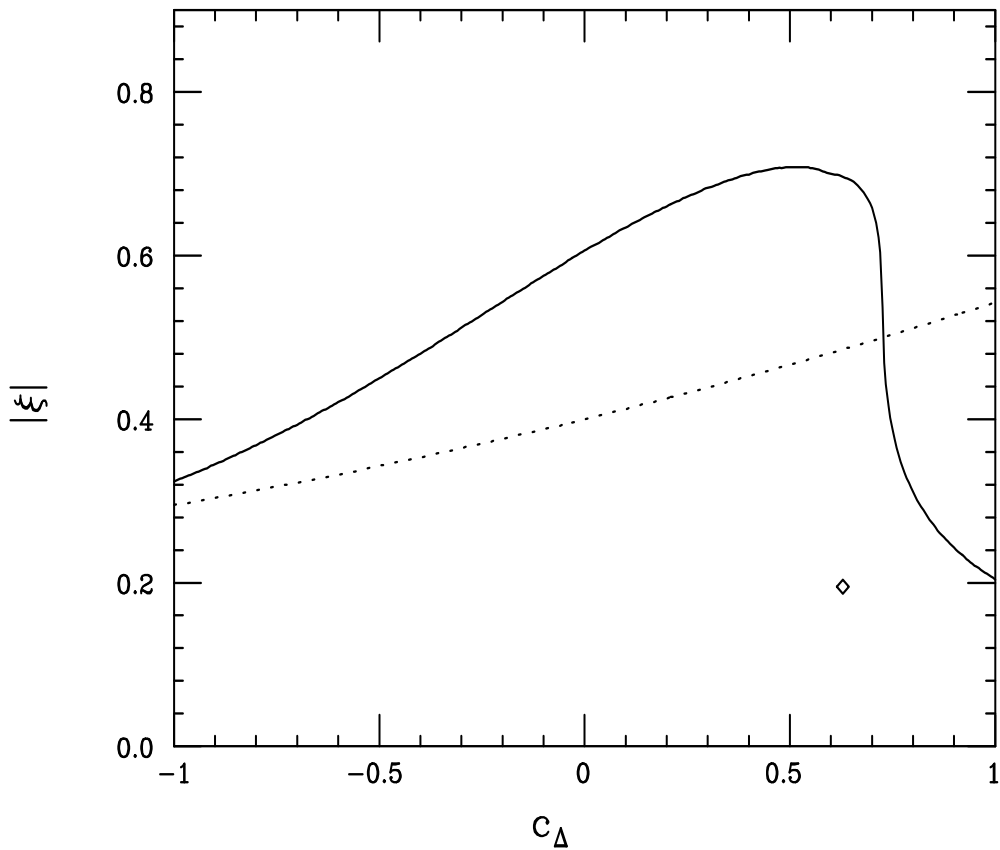,height=14cm,width=16cm,angle=-90}}
\vspace*{-1cm}
\caption{Comparison of the envelope of the $95\%$ CL allowed regions 
obtained with CLEO data(solid curve) with an estimate of the upper bound 
obtainable from the analysis of the 
L3 charged lepton and neutrino data(dotted curve). The location of the 
$\chi^2$ minima from the L3 analysis is also shown.}
\label{l3stuff}
\end{figure}
\vspace*{0.4mm}

\section{The Left-Right Model and $\xi$}

If RH currents do exist, given the CLEO allowed region in the 
$|\xi|-c_\Delta$ plane shown in Fig.1, we want to know if there are any 
sub-regions of this allowed space that can yield a simultaneous lowering of 
the SM predictions for 
$B_\ell$ and $n_c$. To address this question we will need to go beyond the 
physics described by the effective Hamiltonian in Eq.(1) since, \eg, we need 
to discuss non-leptonic decay modes such as $b\to c\bar ud(s)$ and 
$b\to c\bar cs(d)$ and the role RH currents may play in these channels. To do 
this we need to incorporate the physics of 
${\cal H}_{sl}$ into a larger framework, \eg, the LRM{\cite {lrm}}. We remind 
the reader that other frameworks, such as SUSY loops, compositeness or 
$R-$parity violation schemes, are 
also possible{\cite {alex2}} sources of effective RH currents. 

In order to be self-contained let us briefly review the relevant parts of the 
LRM we need for our discussion below; for details of the model the reader is 
referred to {\cite {lrm}}. The LRM is based on the extended gauge group 
$SU(2)_L \times SU(2)_R \times U(1)$. Due to this extension 
there are both new neutral and charged gauge bosons, $Z', W^{\pm}_R$, in 
addition to those present in the Standard Model. In this scenario the 
left-(right-)handed fermions of the SM are assigned to doublets under 
the $SU(2)_{L(R)}$ group and a RH neutrino is introduced. The Higgs fields 
which can directly generate SM fermion masses are thus in {\it bi-doublet} 
representations, \ie, they transform as doublets under both $SU(2)$ groups. 
The LRM is quite robust 
and possesses a large number of free parameters which play an interdependent 
role in the calculation of observables and in the existing constraints on the 
model resulting from various experiments. 

As far as $B$ physics and the subsequent discussion are concerned there are 
several parameters of direct interest. The most obvious free parameter is the 
ratio of the $SU(2)_R$ and $SU(2)_L$ gauge couplings 
~$0.55<\kappa=g_R/g_L\leq 2$; the lower limit is a model constraint while the 
upper one is simply a naturalness assumption. GUT embedding scenarios 
{\it generally} suggest that $\kappa \leq 1${\cite {desh}}. 
For simplicity we will assume that 
$\kappa=1$ in almost all of our discussion below. The extended gauge symmetry 
is broken in two stages. First the $SU(2)_L \times SU(2)_R\times U(1)$ symmetry 
is broken down to the SM via the action of Higgs fields that transform either 
as doublets or triplets under $SU(2)_R$. This choice of Higgs representation 
determines both the mass relationship 
between the $Z'$ and $W_R$ (analogous to the condition that $\rho=1$ in the 
SM with only Higgs doublets and singlets) as well as the nature of neutrino 
masses; in particular, the Higgs triplet 
choice which we employ here allows for the implementation of the 
see-saw mechanism and yields a heavy RH neutrino. 

After complete symmetry breaking the resulting $W_L-W_R$ mixing is described 
by two parameters, a real mixing angle, $\phi$, and a phase, $\omega$. Note 
that it is usually $t=\tan \phi$ which appears in expressions directly 
related to observables. The additional phase, as always, can be a new source 
of CP violation. The mixing between $W_L$ and $W_R$ results in 
the mass eigenstates $W_{1,2}$, with a ratio of masses, $r=M_1^2/M_2^2$, 
(with $M_2 \simeq M_R$). In most models $T$ is then naturally of order a few 
times $r$ or less in the large $M_2$ limit. Of course, $W_1$ is the state 
directly being produced at 
both the Tevatron and LEPII and is identical to the SM $W$ in the $\phi \to 0$ 
limit. We note that when $\phi$ is non-zero, $W_1$ no longer couples to a 
purely LH current. Of course if a heavy RH neutrino is indeed realized then 
the effective {\it leptonic} current coupling to $W_1$ remains purely LH as 
far as all low energy experiments are concerned. 
As is well-known, one of the strongest set of `classical' constraints on 
this model 
arises from polarized $\mu$ decay{\cite {muon}}, which are trivial to satisfy 
in the case of a heavy RH neutrino and this justifies the appearance of only 
LH leptonic couplings in Eq.(1). Removal of these constraints provides 
significantly more freedom in the remaining LRM parameter space. 
Thus the tree-level $\mu$ decay Hamiltonian is just
\begin{equation}
{\cal H}_\mu={g_L^2c_\phi^2(1+rt^2)\over {2M_1^2}}(\bar \nu_{\mu_L}
\gamma_\lambda \mu_L)(\bar e_L \gamma^\lambda \nu_{e_L})\,,
\end{equation}
so that the tree-level definition of $G_F$ is simply
\begin{equation}
{G_F\over {\sqrt 2}}={g_L^2c_\phi^2(1+rt^2)\over {8M_1^2}}\,.
\end{equation}
We see that if $r$ and $t$ are of order $\simeq 10^{-2}$ or less the numerical 
influence of mixing in this relationship will be quite small. 

We note that it is important to be reminded that the extended Higgs sector 
associated with both the breaking of the LRM group down to 
$U(1)_{em}$ and the complete generation of fermion masses may also have an 
important role to play in low energy physics through both the existence of 
complex Yukawa and/or flavor-changing neutral current type couplings. However, 
this sector of the LRM 
is highly model dependent and is of course quite sensitive to the detailed 
nature of the fermion mass generation problem. For purposes of brevity and 
simplicity and because tree-level neutral Higgs exchange can little influence 
the decay processes we are interested in these too will be ignored in the 
following discussion and we will focus solely on the effects associated with 
$W_{1,2}$ exchange. We do note that these additional Higgs fields can 
potentially play a very important role in loop processes as will be briefly 
discussed later. 

Additional parameters arise in the quark sector; in principle the effective 
mass matrices for the SM fermions may be non-hermitian implying that the two 
matrices involved in the bi-unitary transformation needed to diagonalize them 
will be unrelated. This means that the elements of the mixing matrix, 
$V_R$, appearing in the RH charged current for quarks will be {\it unrelated} 
to the corresponding elements of $V_L=V_{CKM}$. $V_R$ will then involve 3 new 
angles as well as 6 additional phases all of which are {\it a priori} 
unknown parameters{\cite {herczeg}}. 
The possibility that $V_L$ and $V_R$ may be unrelated is 
sometimes overlooked when considering the potential impact of the LRM on low 
energy physics and there has been very little detailed exploration of this 
more general situation due to the rather large parameter space. Certainly 
as the elements of $V_R$ are allowed to vary 
the impact of the extended gauge sector on $B$ physics in general 
will be greatly effected. 

Some well-known constraints on the LRM, such as Tevatron direct $W'$ 
searches{\cite {cdfd0}}, are quite sensitive to variations in 
$V_R${\cite {oldt}} as well as the properties of the RH neutrino and 
$W_2$ masses as 
low as $450-500$ GeV can very easily be accommodated by the present data. 
To be conservative, and with future Tevatron 
searches in mind, however, we will assume 
below that $M_2 \geq 720$ GeV{\cite {cdfd0}}, \ie, $r\leq 0.012$, for any $V_R$ 
implying that the magnitude of 
$t$ is also less than $\sim 0.012$. Other constraints on the LRM parameter 
space involve loop processes such as the $K_L-K_S$ mass 
difference{\cite {bbs,lang}} and $b\to s\gamma${\cite {old}}. Clearly the 
bounds obtained from these processes depend not only on the gauge sector but 
also on {\it all} the particles that can participate in the loops such as 
SUSY partners, extra Higgs fields, additional heavy fermions, \etc.,~whose 
existence is sensitive to the finer details of the model. 
These possibilities are beyond the necessities of 
the current discussion where we are solely interested in tree-level $B$ 
decays. Our philosophy as outlined in the introduction will be to leave for 
now all discussions of loop graphs which display any sensitivity to the 
details of the LRM spectrum and take these issues up briefly later. 

Using the definitions above for the LRM parameters we can now express $\xi$ in 
terms of these more fundamental quantities; we find
\begin{equation}
\xi={\kappa t(1-r)\over {(1+rt^2)}}~{e^{i\omega}V_{cb}^R\over {V_{cb}^L}}\,.
\end{equation}
Note that we can absorb the sign of $t$ into the phase $\omega$ here so that 
$t$ can be treated as a positive parameter in our discussion below. As 
mentioned above we will take $\kappa=1$ for simplicity in our numerical 
analysis below; to first order it simply rescales the value of $t$. 
Employing the results from Buras{\cite {buras}} for the value of 
$V_{cb}^{L~inc}(D^*)$  
from inclusive $B$ semileptonic decays, we can invert the expression above 
to obtain 
\begin{equation}
|V_{cb}^R|=(39.9\pm 2.2)\cdot 10^{-3}~{|\xi|\over {[1+|\xi|^2+2\xi c_\Delta
{g\over {f}}]^{1/2}}}~{(1+rt^2)\over {\kappa t(1-r)}}\,,
\end{equation}
so that for typical values such as $|\xi|=0.2$, $c_\Delta$=0, and $x=0.29$, 
we obtain 
\begin{equation}
|V_{cb}^R|=(0.782\pm 0.042)~\Bigg[{10^{-2}\over {\kappa t}}\Bigg]~[1
+{\cal O}(r,rt^2)]\,,
\end{equation}
which suggests that $|V_{cb}^R|$ is reasonably large and perhaps of order 
unity over most of the 
allowed parameter space shown in Fig.~\ref{hqet}. From these considerations 
we learn several 
things which follow immediately from the unitarity of $V_R$: ($i$) A large 
value for $|V_{cb}^R|$ 
implies that the sum $|V_{cd}^R|^2+|V_{cs}^R|^2$ is small thus somewhat 
suppressing potential 
RH contributions to the decays $b\to c\bar cs(d)$, which is fortunate for charm 
counting purposes. If either of these elements were large one might expect a 
significant increase in $n_c$ due to RH current contributions. As we will see 
below, just the opposite occurs. As will be noted, this also assists 
in suppressing RH contributions to $K_L-K_S$ mixing.  
($ii$) Since unitarity requires $|V_{ud}^R|^2+|V_{us}^R|^2+|V_{ub}^R|^2=
|V_{ub}^R|^2+|V_{cb}^R|^2+|V_{tb}^R|^2$ it follows immediately that 
$|V_{ud}^R|^2+|V_{us}^R|^2 >|V_{cb}^R|^2$. However, since $|V_{cb}^R|^2$ is 
apparently large this inequality implies that the sum 
$|V_{ud}^R|^2+|V_{us}^R|^2$ is larger still. This would mean 
that decay modes such as $b\to c\bar ud(s)$ may receive large RH 
contributions. We note that if we further assume that 
$|V_{ud}^R|^2 << |V_{us}^R|^2$ these RH 
contributions may also lead to an increase in $K$ production in $B$ 
decays{\cite {alex}}, which it has been argued is a signal for enhanced 
$b \to sg$. Also if $|V_{ud}^R|^2 << |V_{us}^R|^2$ one finds that the 
Tevatron search reach{\cite {cdfd0}} for $W_2$ would be seriously degraded 
by about a factor of 2{\cite {oldt}} in mass. 
($iii$) It would appear that $|V_{ub}^R|$ will be too small to 
significantly influence $b\to u$ processes, though this needs further 
examination. ($iv$) A large $V_{cb}^R$ implies that the sum $|V_{td}^R|^2
+|V_{ts}^R|^2$ is also large{\cite {alex2}} with implications for the complete 
structure of $V_R$ that we will 
ignore for now but will return to haunt us in our discussion below. ($v$) The 
fact that unitarity requires $|V_{cb}^R| \leq 1$ itself provides an additional 
constraint on the remaining LRM parameters.

\section{Non-leptonic $b\to c$ Decays with RH Currents}

As a final step in our analysis we need the complete non-leptonic Hamiltonian; 
at the tree-level this can now be written down immediately.
For the sample case of $b\to c\bar u d$ we can write, following the notation 
in Refs.{\cite {bsgme,new}},
\begin{equation}
{\cal H}_{nl}={4G_F\over {\sqrt 2}}\Bigg[C_{2L}O_{2L}+C_{12L}O_{12L}+
L\to R ~\Bigg]\,,
\end{equation}
where $O_{2L}=(\bar c\gamma_\mu P_L b)(\bar d\gamma^\mu P_L u)$, 
$O_{12L}=(\bar c\gamma_\mu P_R b)(\bar d \gamma^\mu P_L u)$, \etc.~ and where 
$P_{L,R}$ are helicity projection operators. 
At the {\it weak} scale the operator coefficients are given by
\begin{eqnarray}
C_{2L} &=& (V_{cb}^L)(V_{ud}^{L*}) \,, \nonumber \\
C_{12L} &=& \Bigg[{\kappa t(1-r)\over {(1+rt^2)}}\Bigg](V_{cb}^R)(V_{ud}^{L*}) 
\,, \nonumber \\
C_{12R} &=& \Bigg[{\kappa t(1-r)\over {(1+rt^2)}}\Bigg](V_{cb}^L)(V_{ud}^{R*}) 
\,, \\
C_{2R} &=& \Bigg[{\kappa^2 (r+t^2)\over {(1+rt^2)^2}}\Bigg](V_{cb}^R)
(V_{ud}^{R*}) \,. \nonumber 
\end{eqnarray}
Note that if we neglect the light quark masses the appropriate phase space 
functions for this particular decay mode will be given by $f$ and $g$. 
The modifications necessary for the study of the decay $b\to c\bar u s$ are 
obvious. Similarly for the corresponding 
decays $b\to c\bar cs(d)$ we simply change the appropriate CKM factors in 
the above and employ the appropriate phase space functions, $f_c$ and $g_c$, 
which are given by the phase space integrals $I_{1,2}$ in Section 2 with the 
replacement $y\to x$. For $x=0.29$ these are found numerically to be 
$f_c \simeq 0.222$ and $g_c \simeq -0.086$. The neglect of the strange quark 
mass, $m_s \simeq 100-150 ~MeV$, is found to be an excellent approximation 
here. 

To proceed with this calculation we need to compute the QCD corrections 
associated with the Renormalization Group running from the weak matching scale 
down to $\mu \sim m_b$. To this end we follow the analysis of 
Bagan \etal{\cite {bagan}} which allows us to write the partial width for 
this process as 
\begin{equation}
\Gamma(b\to c\bar ud(s))= \Gamma_{SM}\Bigg[1+\eta_1+\eta_2+\eta_3\Bigg]\,,
\end{equation}
where $\Gamma_{SM}=3X_1\Gamma_0f|V_{cb}^L|^2(|V_{ud}^L|^2+|V_{us}^L|^2)$, 
$\Gamma_0$ is the canonical $\mu$ decay width with the replacement 
$\mu \to m_b$, and 
$X_1$ represents the results of SM QCD corrections(to which we will return 
below). The $\eta_i$ are LRM contributions which given by
\begin{eqnarray}
\eta_1 &=& \Bigg[{\kappa(r+t^2)\over {t(1-r)}}\Bigg]^2|\xi|^2y\,, \nonumber \\
\eta_2 &=& {X_2\over {X_1}}\Bigg[|\xi|^2+{\kappa^2 t^2(1-r)^2\over {(1+rt^2)^2}
}y\Bigg]\,, \\
\eta_3 &=& 2{g\over {f}}{X_3\over {X_1}}Re(\xi)\Bigg[1+{\kappa^2(r+t^2)\over {
(1+rt^2)}}y\Bigg]\,, \nonumber 
\end{eqnarray}
with $y=(|V_{ud}^R|^2+|V_{us}^R|^2)/(|V_{ud}^L|^2+|V_{us}^L|^2)\simeq
~(|V_{ud}^R|^2+|V_{us}^R|^2)$. As pointed out in the discussion above, 
if $|V_{cb}^R|$ 
is large we anticipate that $y$ is near unity. For the decays 
$b\to c\bar cs(d)$ we make the obvious CKM replacements and the change 
the $X_i \to X_i'$, $f,g\to f_c,g_c$ and let $y\to y_c$ where 
$y_c=(|V_{cd}^R|^2+|V_{cs}^R|^2)/(|V_{cd}^L|^2+|V_{cs}^L|^2)\simeq
~1-|V_{cb}^R|^2$, with the last near equality resulting from unitarity and 
the fact that $|V_{ub}^L|^2$ is very small. If $|V_{cb}^R|^2$ is large then 
clearly $y_c$ must then be small. 

At leading order(LO) in QCD the $X_i=X_i'$ are completely calculable and are 
simple polynomials in the parameter
\begin{equation}
z=\Bigg[{\alpha_s(M_W)\over {\alpha_s(\mu)}}\Bigg]^{3/23}\,,
\end{equation}
and its inverse; here we will assume that $\alpha_s(M_Z)=0.118$ and 
$\mu \sim m_b$. Explicitly, we obtain 
\begin{eqnarray}
X_1 &=& {1\over {3}}\Bigg[2z^4+z^{-8}\Bigg]\,, \nonumber \\
X_2 &=& {1\over {9}}\Bigg[8z^2+z^{-16}\Bigg]\,, \\
X_3 &=& {1\over {9}}\Bigg[4z^3+4z^{-3}+2z^{-6}-z^{-12}\Bigg]\,, \nonumber 
\end{eqnarray}
where we have made use of the results of Altarelli and Maiani{\cite {hokim}} 
as well as Cho and Misiak{\cite {old}}. 
Note all $X_i\to 1$ as $z\to 1$ and the QCD corrections vanish. In the SM, NLO 
multiplicative corrections to the LO values of $X_1$ and $X_1'$ are now 
known{\cite {bagan}} to be $\simeq 1.061$ and $\simeq 1.29$, respectively, 
for $\mu=m_b$, $x=0.29$, and using pole quark masses($m_b=4.8$ GeV), both 
of which we 
adopt in the numerical analysis below. Unfortunately, the corresponding NLO 
corrections to  $X_{2,3}$ and $X_{2,3}'$ are 
not yet known. The best we can do until such calculations are performed is to 
follow Voloshin's philosophy and assume the multiplicative corrections in 
these cases are essentially the same as those for $X_1$ and $X_1'$. Since, as 
we will see below, we will be more interested in the {\it shifts} in the 
values of $n_c$ and $B_\ell$ due to RH currents than the values themselves, 
we anticipate that this assumption may be a fair approximation. We note that 
in making this assumption we are also ignoring the possibility that the 
detailed LRM particle spectrum may lead to substantial modifications in these 
SM values, in particular, those contributions arising from penguins. These 
assumptions need to be verified by future direct calculations.

\section{$\delta B_\ell$ and $\delta n_c$}

From the discussion in the previous section we are ready to calculate both 
$\delta B_\ell=B_\ell(LRM)-B_\ell(SM)$ and $\delta n_c=n_c(LRM)-n_c(SM)$ 
where the SM results are given by the above expressions in the limit where 
all RH couplings are turned off. As is well known, the combined experimental 
and theoretical situation is quite puzzling. From the reviews of both Drell 
and Sachrajda{\cite {prob}} we see that $B_\ell=0.1018\pm 0.0040$ on the 
$\Upsilon (4S)$ while $B_\ell=0.1095\pm 0.0032$ at the $Z$. Similarly, 
$n_c=1.119\pm 0.053$ and $1.202\pm 0.067$ on the $\Upsilon (4S)$ and $Z$, 
respectively. Numerically, in the SM limit our calculations essentially 
reproduce the earlier results of Bagan \etal{\cite {bagan}} which we have 
closely followed; in this limit we 
obtain $B_\ell=0.123$ and $n_c=1.24$ for the SM predictions assuming $x=0.29$ 
and $\mu=m_b$. We will 
implicitly assume that there are no new $b\to no~charm$ final states, such as 
$b\to sg$, which are enhanced due to RH currents.  It is clear that if we 
take these experimental 
numbers at face value we would like to decrease the theoretical 
predictions for $B_\ell$ by $0.015-0.020$ and $n_c$ by at least 0.03.

Our analysis consists of 
an extensive scan of the model parameter space spanned by $r$, $t$, 
$|\xi|$, $c_\Delta$ and $y$ and demanding that a number of requirements be 
satisfied simultaneously.
Our input parameters are chosen as follows. We begin by picking a `point' 
inside of the CLEO allowed region in the $c_\Delta-|\xi|$ plane so that 
this constraint is already satisfied. We assume the scale size of the 
$c_\Delta-|\xi|$ grid to be $0.01\times 0.01$ so that are approximately 
$1.5\cdot 10^4$ points in this sample. 
Next, we choose a value for the two LRM parameters $r$ and $t$; for simplicity 
$\kappa$ is set to unity. Keeping in mind the CDF/D0{\cite {cdfd0}} bounds and 
the strong suggestion that $t$ cannot be much larger than $r$, we let 
$r=0.0025$, 0.005, 0.0075, 0.010 or 0.012 and allow $t$ to vary over the range 
0 to 0.012 in steps of 0.0005. (Remember that due to the phase freedom in 
angle $\Delta$ we can treat $t \geq 0$ in this discussion.)
Clearly if the $W_2$ mass is too large and/or the 
mixing angle is too small the effects of RH currents will not be of a 
noticeable magnitude. This restricts our attention to $W_2$ masses in the 
approximate range $730-1600$ GeV. 
Thus we see that for every choice of $c_\Delta$ and $|\xi|$ there are 120 
pairs of $(r,t)$ values giving us a total of $\simeq 1.8\cdot 10^6$ points to 
examine in the $r-t-|\xi|-c_\Delta$ parameter subspace. 

The first constraint we impose is the requirement that $|V_{cb}^R|$ be less 
than unity by using Eq. (32). Of course if this 
constraint is not satisfied for any of the $r$ or $t$ values this point on the 
$c_\Delta-|\xi|$ grid is removed from any further consideration. If satisfied, 
the result fixes the value of $y_c$ in the subsequent calculations. 
To proceed we must choose a value of $y$ in the range $0<y<1$ which 
we do in grid steps of 0.01. We then impose our second constraint that 
$y\geq |V_{cb}^R|^2$ so that only the larger of the $y$ values survive. Out of 
the original $\simeq 1.8\cdot 10^8$ points in the five-dimensional 
$r-t-|\xi|-c_\Delta-y$ 
parameter space being scanned, only $\simeq 27.5\cdot 10^6$ survive these first 
two constraints.  

For these remaining points we next calculate $\delta B_\ell$ and 
$\delta n_c$ for each particular choice of input parameters and impose our 
final loose requirement that $\delta B_\ell \leq -0.01$ and 
$\delta n_c \leq -0.025$. Again, if these constraints cannot be met at a 
particular point on the $c_\Delta-|\xi|$ grid, independently of the chosen 
values of $r,t$ and $y$, it is removed. 
Only 6284 points in the the $r-t-|\xi|-c_\Delta-y$ five-dimensional 
parameter space now remain; this number is further reduced to 972 if we 
strengthen our requirement on $\delta n_c$ to be $\leq -0.03$. 
It is clear from these numbers that a rather high degree of fine-tuning is 
required to push $B_\ell$ and $n_c$ in the proper direction and to produce 
shifts with interesting magnitudes. For most values of the parameters the 
resulting shifts in $B_\ell$ and/or $n_c$ are much too small to be of 
interest. The combination of these requirements is found to be extremely 
demanding on the model parameter space, yet two distinct 
sub-regions do survive.

\vspace*{-0.5cm}
\nn
\begin{figure}[htbp]
\centerline{
\psfig{figure=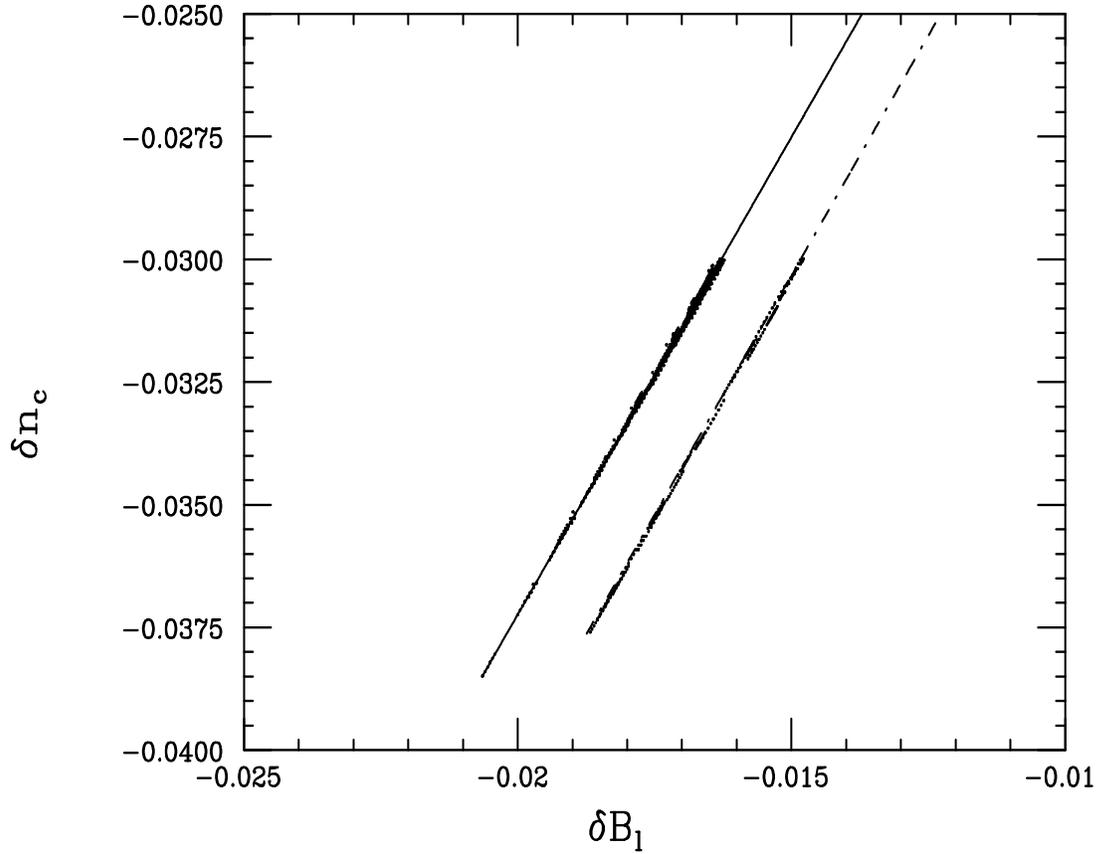,height=14cm,width=16cm,angle=-90}}
\vspace*{-1cm}
\caption{Location of surviving points in the $\delta B_\ell-\delta n_c$ plane.  
The 972 survivors of the $\delta n_c <-0.03$ cut are shown explicitly. 
The lines represent smoothed versions of the actual locations. The 
solid(dash-dotted) line corresponds to the solutions with $c_\Delta >(<) 0$. }
\label{res}
\end{figure}
\vspace*{0.4mm}

If one plots the values of $\delta B_\ell$ and $\delta n_c$ for the survivors 
we find that they essentially lie only along two straight lines in the 
$\delta B_\ell-\delta n_c$ plane with the choice of line depending upon the 
sign of $c_\Delta$ as shown in Fig.~\ref{res}. The corresponding location of 
these same points with $\delta n_c \leq -0.03$ projected onto the 
$c_\Delta-|\xi|$ plane are shown in Fig.~\ref{res2}. It is amusing to note 
that the points with $c_\Delta>0$ lie within the region associated with the 
fit to CLEO's data on the $\chi$ distribution in $B\to D^*(\to D\pi)\ell \nu$ 
obtained above. 
Assuming $\delta n_c \leq -0.025(-0.03)$ approximately $92.5(75.1)\%$ 
of the survivors are found to lie in 
the $c_\Delta >0$ region. The fractional volume of the $\delta n_c \leq -0.025$ 
parameter space which also allows $\delta n_c \leq -0.03$ is $\simeq 15.5\%$. 
While the $c_\Delta<0$ parameter space is only reduced to $51.4\%$ of its 
previous size by strengthening this $\delta n_c$ cut, the $c_\Delta>0$ 
subspace is drastically reduced to only $12.6\%$ of its previous population by 
this same cut. 

What are some of the various properties of the parameter space points that 
satisfy all our requirements? Mostly they are exactly what one would naively 
expect. First, all of the 972 survivors have $t\geq 0.0095$ since larger mixing 
angles are required to enhance the contributions of the RH currents. 
Second, in all cases $|V_{cb}^R|\geq 0.908$ and there is a significant 
preference for larger values of $r$, \ie, there are only 4(37) cases with 
$r=0.0025(0.005)$.

\vspace*{-0.5cm}
\nn
\begin{figure}[htbp]
\centerline{
\psfig{figure=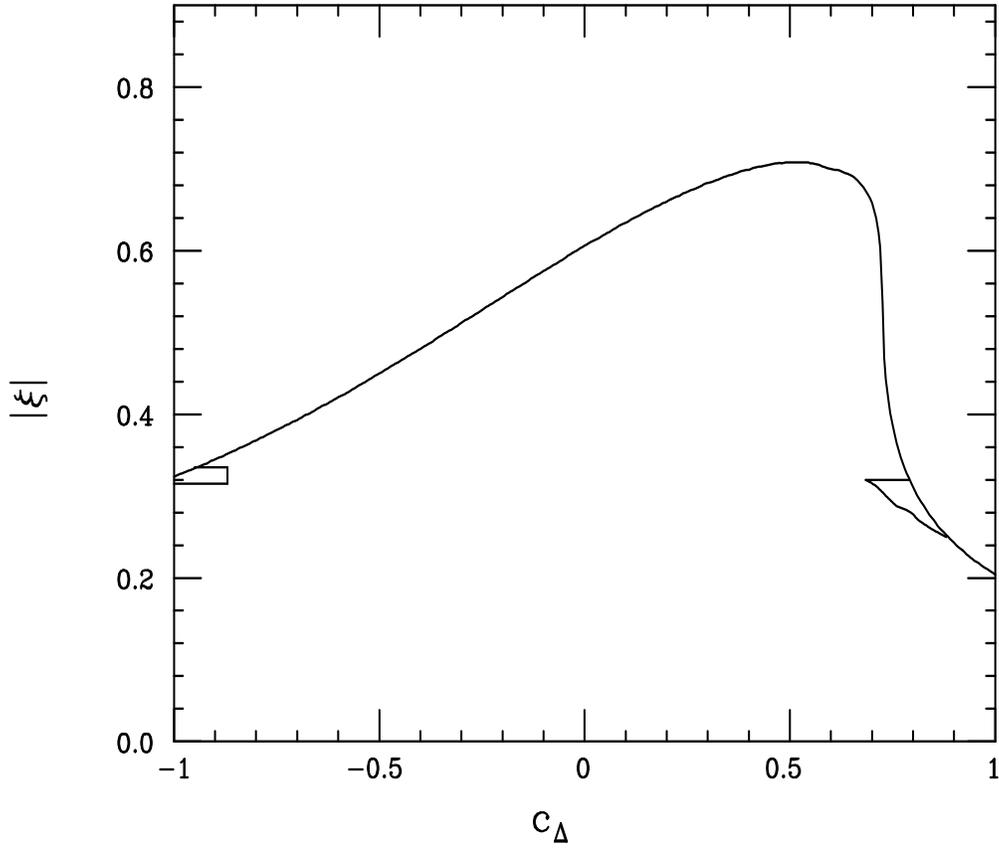,height=14cm,width=16cm,angle=-90}}
\vspace*{-1cm}
\caption{Locations of the zones containing the 972 surviving points in the 
$c_\Delta-|\xi|$ plane, in comparison to the envelope of that allowed by CLEO 
at $95\%$ CL, which simultaneously 
satisfy $\delta n_c \leq -0.03$ and $\delta B_\ell \leq -0.01$.}
\label{res2}
\end{figure}
\vspace*{0.4mm}

\section{Discussion and Conclusions}

The chirality of the $b\to c$ coupling is one of the most important quantities 
in $B$ physics. The original work of Gronau and Wakaizumi demonstrated to us 
just how little was actually known about this coupling at that time. Since 
then, after extensive theoretical and experimental effort, the situation 
remains far from being completely clarified. 
While CLEO and ALEPH have certainly demonstrated that the $b\to c$ 
coupling is dominantly LH in agreement with the SM, their results remain 
consistent with the possibility of a sizeable RH coupling. Furthermore, the 
interpretation of the low value of the $\Lambda_b$ polarization obtained by 
L3 remains ambiguous and could either be a first signal 
for RH currents or simply a sign of our ignorance of the strong interactions.  
All of these experimental analyses have been based on relatively small sample 
sizes and need to be repeated and improved upon.

As we saw in the analysis above, the exclusive $B\to D^*$ semileptonic decay 
provides us with a large number of observables that can be used to probe for 
RH couplings of reasonably small strength. In addition to the overall 
partial width, expressible in terms of $V_{cb}^{L~exc}$, the measurements of 
the $\cos \theta_\ell$ and $\cos \theta_V$ distributions lead directly to 
the quantities $A_{FB}$ and $\Gamma_L/\Gamma_T$, respectively. By using HQET 
we performed a fit to the present CLEO results for these quantities and 
demonstrated that the current bound on the RH coupling strength still remains 
rather poor especially if it is allowed to be complex. Improved statistics 
available at upcoming $B$ factories will help tremendously here. Furthermore, 
while we showed that the $q^2$ distribution was not very sensitive to RH 
couplings, the $\chi$ distribution was found to be particularly so and 
yielded tantalizing indications for the existence of RH currents. Present 
CLEO data was shown to indicate that future measurements of this distribution 
will be extremely useful in either constraining or discovering RH couplings. 
More recent but yet unpublished CLEO results{\cite {unp1}} seem to strengthen 
the case for the existence of right-handed currents based on the $\chi$ 
distribution. 

The low $\Lambda_b$ polarization result obtained by ALEPH also remains 
tantalizing and 
certainly needs updating. Unfortunately, given our incomplete knowledge of 
QCD, any interpretation of the result in terms of RH 
currents can not be made at present. However, if high precision measurements 
of the lepton and missing energy spectra become available with only a factor 
of a few increase in statistics, we saw in the 
analysis above that sufficient observables do exist to separate the two 
possible explanations. Further observables may be found to strengthen any 
conclusions one may draw from future data. The low $\Lambda_b$ measurement 
seems to be confirmed by as yet unpublished data from both ALEPH and 
DELPHI{\cite {unp2}}.

Under the assumption that $b\to c$ RH currents {\it do} exist consistent 
with the 
bounds from CLEO, we have tried to address the question raised by Voloshin as 
to whether such new interactions could assist in solving the long standing 
problem associated with $B_\ell$ and $n_c$. To address this point we needed 
to go beyond the model independent results of the previous section and 
incorporate our $b\to c$ RH coupling scenario into a larger framework, the 
most natural one being the LRM. Within this scheme, making a number of 
assumptions about both the detailed particle spectrum of the model and the 
nature of the NLO QCD corrections to the RH pieces of the nonleptonic 
Hamiltonian operator coefficients, we were able to demonstrate that two 
small regions of the LRM parameter space do exist that push both $B_\ell$ 
and $n_c$ in the right directions with sufficient magnitudes to be 
phenomenologically interesting. These small parameter space regions result 
from a reasonably highly tuned set of LRM parameters and in all cases 
$V_{cb}^R$ was found to have a magnitude of order unity.

Hopefully measurements at the new $B$ factories which are soon to turn on 
will yield signals for physics beyond the Standard Model. Perhaps right-handed 
currents will be among them.

\noindent{\Large\bf Acknowledgements}

The author would like to thank J.L. Hewett, A. Kagan, C. Diaconu, S. Stone, 
Y. Grossman, M. Dittmar, A. Ryd, K. Kiers, J. Wells and M. Worah for 
discussions related to this work.

\newpage

\vspace{1in}
\centerline{\bf APPENDIX}
\vspace{0.1in}

In this Appendix, we outline some of the implications of the scenarios 
discussed above that lead to lower values of both $B_\ell$ and $n-c$ while 
satisfying the CLEO constraints. Such results, for example, may lead one 
to speculate on just what form the matrix $V_R$ might take if this type of 
solution to the $B_\ell-n_c$ problem were to be realized. This will directly 
lead to a number of wide ranging implications in all low energy sectors of 
the theory and not just in $B$ physics. (In fact, there are too many for us to 
comment upon here with any depth of discussion.) Unfortunately, 
we do not yet have available a global analysis of RH current phenomenology for 
arbitrary forms of $V_R$ with a left-right mixing at the per cent level. Such 
an analysis would be extremely useful for our discussion but is far beyond 
the scope of the present paper. 

If we hypothesize{\cite {lang,new}} that in each row or column there is a 
single element with a magnitude near unity, as is true for the conventional CKM 
matrix, then there are only two RH mixing matrices which allow for large 
$V_{cb}^R$. Following the notation employed in our earlier work{\cite {new}}, 
we can write these `large element' forms symbolically, neglecting any phases, 
as 
\begin{eqnarray}
{\cal M}_C & =  \left( \begin{array}{ccc}
                         1 & 0 & 0 \\
                         0 & 0 & 1 \\
                         0 & 1 & 0
                         \end{array}\right)\,,\quad\quad
{\cal M}_D & =  \left( \begin{array}{ccc}
                         0 & 1 & 0 \\
                         0 & 0 & 1 \\
                         1 & 0 & 0
                         \end{array}\right)\,,
\end{eqnarray}
with the true $V_R$ being a perturbation about one of these skeletons, just 
as the CKM is a perturbation about the diagonal unit matrix. As noted 
elsewhere{\cite {new}} the structure of these matrices combined with small 
values of $t$ allow us to easily circumvent the traditional constraints 
on the LRM from the magnitude of $K_L-K_S$ mixing. However, we also observe  
here the necessity that at least one of $V_{td}^R$ or $V_{ts}^R$ is large 
which can lead us to some potential problems with the observed rate for and 
limits upon the processes $b\to d,s\gamma${\cite {alex2}}. As has been 
discussed in previous analyses of the $b\to s\gamma$ process within the 
context of the LRM{\cite {old,bsgme,new}}, interference terms between the 
RH and LH 
$W_1-W_2$ contributions obtain enhancements by helicity-flip factors of 
$m_t/m_b$ though 
they are also simultaneously suppressed by a factor of $t$. While the pure 
SM piece is proportional to the product $V_{tb}^LV_{td,s}^L$, these new 
interference terms are correspondingly 
proportional to either $V_{tb}^LV_{td,s}^R$ or $V_{tb}^RV_{td,s}^L$; the 
later being quite small in our case. However, here we have seen that at 
least one of the products $V_{tb}^LV_{td,s}^R$ is of order unity. This 
implies that in the models we have found here such LH-RH interference 
terms arising from $W_{1,2}-t$ quark 
penguins may be dangerously large, by factors of order 10, in at least one of 
the $b\to s\gamma$ or $b\to d\gamma$ modes. Of course one may argue that we are 
most likely quite ignorant of the true LRM model spectrum and that loop 
contributions from the non-$W_i-t$ diagrams may eliminate this problem. As is 
well known, SUSY and charged Higgs exchanges can, for example, yield 
significant contributions to these penguins and can possibly leading to a 
fine-tuned cancellation amongst the various pieces. This is an unnatural, yet 
potentially possible, solution.

Another potentially important constraint arises from the determination of 
the relative branching fractions for the $B\to \psi \pi$ and $B\to \psi K$ 
modes by CLEO{\cite {sheldon}} to be $4.3\pm 2.3\%$. This measurement was 
inconsistent with the predictions of the Gronau and Wakaizumi model which 
predicted a ratio of $\sim 10^{-7}$. For the class of models presented here, 
this result implies only that $V_{cd}^R$ is probably somewhat smaller than 
$V_{cs}^R$, which is not an unexpected result. 

As is well-known there are many other non$-B$ physics 
constraints on the form of $V_R$ which need to be examined. These are mostly 
concerned with the specific elements $V_{ud,s}^R$; 
some of these have a rather long, even controversial, history. 
Many of these constraints have been extensively reviewed in detail some time 
ago by Langacker and Uma Sankar{\cite {lang}} and as stated above 
it is beyond the scope of the 
present paper to discuss them at any length except for several comments. 
These low-energy constraints include, amongst others, potential violations of 
CKM universality, as suggested by Wolfenstein{\cite {wolf}}, and/or violations 
of PCAC relations, as suggested by Donoghue and Holstein{\cite {don}}.   
For the universality constraint, we note that Buras{\cite{buras}} 
reports $\sum_i|V_{ui}^{L~eff}|^2=0.9972\pm 0.0013$, which is more than 
$2\sigma$ below the SM expectation, perhaps hinting at new physics. {\it If} 
no other new physics sources enter other than the existence of $V_R$, we 
can use the result of this sum to constrain both $Re(V_{ud,s}^R)$ even when 
$\kappa t \sim 0.01$. For example, using $|V_{ud}^L|\sim 0.98$ and 
$|V_{us}^L/V_{ud}^L|\sim 0.22$, this constraint implies 
\begin{equation}
-0.133\pm 0.066 \simeq \Bigg[{{\kappa t}\over {10^{-2}}}\Bigg]~[|V_{ud}^R|
\cos \Delta_d +0.22|V_{us}^R|\cos \Delta_s]\,,
\end{equation}
where $\Delta_i$ is sum of $\omega$ plus the phase of $V_{ui}^R$. This 
constraint is easily satisfied for either of the two forms of $V_R$ suggested 
above assuming reasonable $\Delta_i$. Some possibilities, suggested by an 
earlier analysis of Matrix D{\cite {new}} are to either 
have $V_{us}^R$ essentially unit magnitude but with a rather large 
phase together with 
$|V_{ud}^R|\sim \lambda^2 \simeq 0.05$ with arbitrary phase or to have instead 
$|V_{ud}^R|\sim \lambda \simeq 0.2$ with bot $V_{ud,s}^R$ having sizeable 
phases. 
(At this point we remind the reader, however, that in extended gauge theories 
such as the LRM there can be other potentially significant contributions to 
universality violation, \eg, $Z'$ exchange, as has been discussed by Marciano 
and Sirlin{\cite {bill}}.) 
Interestingly, such possible solutions are also found to easily 
satisfy a number of additional constraints including those from  
PCAC{\cite {don}} (though these need to be updated), those from muon capture on 
$^3$He{\cite {gova}}, and those on the phase of $g_A/g_V$ in neutron beta 
decay{\cite {pdg}}. Similarly, the scaling of the strengths of the $V,A$ 
currents imply that the extracted value of the ratio $(g_A/g_V)/f_\pi$ is 
exactly the same as in the SM with no violation of the Goldberger-Treiman 
relation{\cite {gold}} occurring in the presence of RH currents. While safely 
avoiding all these bounds, however, these solutions do not help in explaining a 
possible disparity between the values of $V_{ud}^L$ extracted from neutron 
decay and that obtained from $0^+\to 0^+$ and $^{19}$Ne beta 
decay{\cite {hag}}. This at the very least would require a quite sizeable 
$V_{ud}^R$. 

In these same scenarios one might expect somewhat larger effects due to RH 
currents to now appear in the strange quark sector. Perhaps one of the most 
significant effects of RH currents here, apart from overall changes in 
normalizations of constants, is in the $F$ and $D$ parameters describing 
hyperon decay. The values extracted for these parameters from $\Delta S=0$ 
and $\Delta S=1$ transitions, corrected for the not yet completely understood 
$SU(3)$ breaking effects, would appear somewhat different. The reason here is 
clear: the ratio of the axial-vector to vector coupling constants in the two 
cases are shifted away from their SM values by different amounts depending on 
the form of $V_R$. Although the data remains rather poor, this possibility is 
not unsupported by the recent analysis of Ratcliffe{\cite {pgr}}. The 
implications are, of course, far reaching and extend as far as tests of the 
Bjorken Sum Rule{\cite {bj}}. We also remind the reader of the well 
known{\cite {pdg}} potential discrepancy between the value of $V_{us}$ 
extracted from the vector current coupling in $K_{e3}$ decays and that from 
hyperon decay data, which probes both axial-vector as well as vector couplings. 

Clearly, is a possible signature of RH currents arises in $B$ decays, the 
search for their influence elsewhere becomes ever more important.

\newpage

%
\def\MPL #1 #2 #3 {Mod. Phys. Lett. {\bf#1},\ #2 (#3)}
\def\NPB #1 #2 #3 {Nucl. Phys. {\bf#1},\ #2 (#3)}
\def\PLB #1 #2 #3 {Phys. Lett. {\bf#1},\ #2 (#3)}
\def\PR #1 #2 #3 {Phys. Rep. {\bf#1},\ #2 (#3)}
\def\PRD #1 #2 #3 {Phys. Rev. {\bf#1},\ #2 (#3)}
\def\PRL #1 #2 #3 {Phys. Rev. Lett. {\bf#1},\ #2 (#3)}
\def\RMP #1 #2 #3 {Rev. Mod. Phys. {\bf#1},\ #2 (#3)}
\def\ZPC #1 #2 #3 {Z. Phys. {\bf#1},\ #2 (#3)}
\def\IJMP #1 #2 #3 {Int. J. Mod. Phys. {\bf#1},\ #2 (#3)}

\end{document}